# Intrinsic Magnetic Topological Materials

Yuan Wang*, Fayuan Zhang*, Meng Zeng*, Hongyi Sun*, Zhanyang, Hao, Yongqing Cai, Hongtao Rong, Chengcheng Zhang, Cai Liu, Xiaoming Ma, Le Wang, Shu Guo, Junhao Lin, Qihang Liu, Chang Liu[#] and Chaoyu Chen[#]

Shenzhen Institute for Quantum Science and Engineering (SIQSE) and Department of Physics, Southern University of Science and Technology (SUSTech), Shenzhen 518055, China.

* Equal contribution

[#] Correspondence should be addressed to C.L. (liuc@sustech.edu.cn) and C.C. (chency@sustech.edu.cn)

**ABSTRACT**

Topological states of matter possess bulk electronic structures categorized by topological invariants and edge/surface states due to the bulk–boundary correspondence. Topological materials hold great potential in the development of dissipationless spintronics, information storage and quantum computation, particularly if combined with magnetic order intrinsically or extrinsically. Here, we review the recent progress in the exploration of intrinsic magnetic topological materials, including but not limited to magnetic topological insulators, magnetic topological metals, and magnetic Weyl semimetals. We pay special attention to their characteristic band features such as the gap of topological surface state, gapped Dirac cone induced by magnetization (either bulk or surface), Weyl nodal point/line and Fermi arc, as well as the exotic transport responses resulting from such band features. We conclude with a brief envision for experimental explorations of new physics or effects by incorporating other orders in intrinsic magnetic topological materials.

**Keywords:** Intrinsic magnetic topological insulator, Magnetic topological metals, Magnetic Weyl semimetal, Topological surface states, Magnetic gap.

## 1. Introduction

The first two decades of the new millennium have witnessed the surge of topological states of matter, whose electronic structures can be categorized by topological invariants. The prediction [1-6] and realization [7-11] of two-dimensional (2D) and three-dimensional (3D) topological insulators (TIs) have caused a paradigm shift to predict, understand and make use of quantum materials based on the topology of their band structures. Started with the 2D TI, quantum spin Hall (QSH) effect realized based on HgTe/CdTe quantum well [6,7] has revealed to the world the fundamental novelty and potential application of topological materials. QSH state is insulating in the bulk but has a pair of one-dimensional (1D) conducting edge states protected by time-reversal symmetry. Electrons in the 1D edge states move without elastic backscattering by nonmagnetic impurities, holding potential for dissipationless spintronics. Likely, a 3D strong TI is also insulating in the bulk and has 2D gapless topological surface states (TSSs). 3D TI was first realized based on Bi-Sb alloys [8] and then on $Bi_2Se_3$ family [5,9-11]. The robustness of topological protection to the TSSs from nonmagnetic perturbations has been experimentally demonstrated [12-14], pointing to feasible electronic and spintronic applications. Importantly, based on magnetically doped $Bi_2Te_3$ films, quantum anomalous Hall (QAH) effect was realized [15,16], a milestone towards low-power-consumption electronics without the need for applied magnetic field.

Besides insulators, quantum materials can also be metals and semimetals according to the detailed band structure around the Fermi level. After TIs, topological semimetals emerged as novel states of matter with degenerate band crossing close to which the band dispersion can be described by the massless 3D Weyl and Dirac equations [17-23]. In a Dirac semimetal (DSM), the conduction and valence bands touch at discrete (Dirac) points with linear dispersion, forming bulk (3D) Dirac fermions. Given broken time-reversal or inversion symmetry, 3D Dirac fermion can be separated in the momentum space into two Weyl fermions (chiral massless fermions as a description of neutrinos with neglected mass in high-energy physics), resulting into topological Weyl semimetals (WSMs). Such nontrivial electronic features could bring into novel electrical and thermal transport behaviors such as anomalous Hall effect, anomalous Nernst effect, chiral anomaly signified by negative magnetoresistance and non-saturating magnetoresistance (see Refs [17,24-26] for comprehensive reviews). Such properties come from the enhanced Berry curvature hosted in the Dirac/Weyl type band structure, which exhibits extreme responses to external stimuli such as magnetic field, voltage or current bias, temperature gradient and optical excitation.

There have been hundreds of materials predicted as 3D strong TI [27-29] and dozens of them have been experimentally verified, usually through direct observation of their TSS Dirac cones by angle-resolved photoemission spectroscopy (ARPES) [30-32]. By comparison, magnetic TIs, especially intrinsic magnetic TIs, are limited in the material candidates. So far there is only $(MnBi_2Te_4) \cdot (Bi_2Te_3)_n$ ($n = 0, 1, 2, 3$) family which has been intensively studied as an intrinsic magnetic TI [33-39]. There are also many materials predicted and demonstrated as DSMs and WSMs, most of which are time-reversal invariant and only few materials have been studied as magnetic WSMs [17, 24-26]. Recently, there appeared several layered material families with hexagonal/Kagome lattices which host Dirac cones gapped by ferromagnetic (FM)/antiferromagnetic (AFM) order, such as $Fe_3Sn_2$ family [40]. In the 2D limit, these systems with gapped Dirac cones can be viewed as Chern insulating phase with quantized anomalous Hall conductance [41], given the Fermi level is positioned in the Dirac gap. In this sense, these materials share the same topological characters (the Chern number $C$) as intrinsic magnetic TIs. However, for 3D materials, the band structure is complicated by the coexisting trivial bands which locate at the same energy region as the Dirac gap, rendering such materials in metal phase with coexisting trivial and nontrivial conduction. Consequently, we feel it more appropriate to term such materials as magnetic topological metals. While quantized transport response from the edge conduction can be realized in intrinsic magnetic TIs by tuning the Fermi level in the gap of both bulk and surface bands, there is always transport contribution from the trivial bands in magnetic topological metals no matter where the Fermi level is. It is noted that there is no strict theoretical picture describing topological metals since the metallicity doesn't come from band topology but trivial bands. We choose this term only to emphasize its distinction from intrinsic magnetic TIs and topological SMs.

In this review, we focus on the recent progress in the exploration of these various kinds of intrinsic magnetic topological materials, categorized mainly into three groups: intrinsic magnetic topological insulators, magnetic Weyl/Dirac semimetals and other magnetic topological metals. We will present representative materials for these novel topological states of matter, pay special attention to their characteristic band features such as the gap of topological surface state Dirac cone, gapped bulk Dirac cone, Weyl nodal point/line and Fermi arc, as well as the exotic transport responses resulting from such band features. There are also other intrinsic magnetic topological states of matter which have been proposed theoretically, yet lacking affirmative experimental evidence, such as topological Möbius insulators [42-44]. We briefly discuss the opportunities to explore new states of matter and novel physical properties based on intrinsic magnetic topological materials.

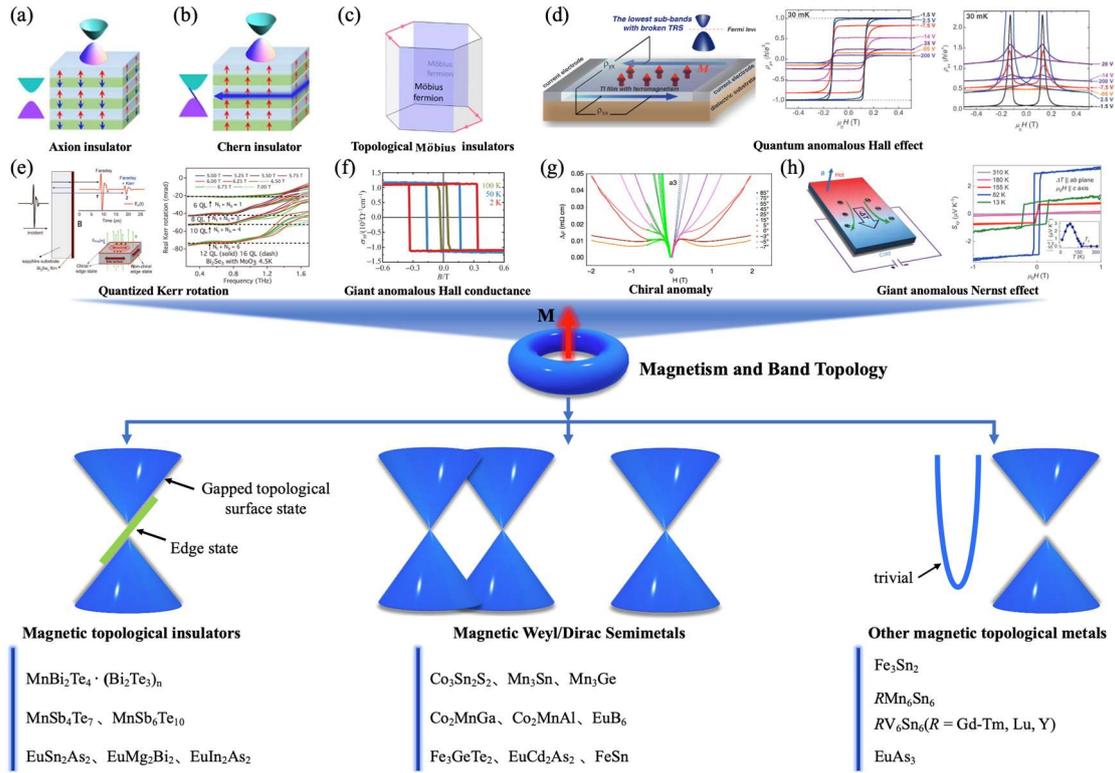

FIG. 1. Family tree of intrinsic magnetic topological materials. The combination of intrinsic magnetism and band topology gives birth to exotic states of matter such as axion insulator (a, adapted from [45]), Chern insulator (b, adapted from [45]), topological Möbius insulator (c, adapted from [44]) and so on, manifesting novel properties such as quantum anomalous Hall effect (d, adapted from [46]), quantized magneto-optical effect (Kerr rotation, e, adapted from [47]), giant anomalous Hall conductance (f, adapted from [48]), chiral anomaly (g, adapted from [49]), giant anomalous Nernst effect (h, adapted from [50,51]) and so on. Magnetic topological materials can be roughly categorized into magnetic topological insulators, magnetic Weyl/Dirac semimetals and other magnetic topological metals, with each of these states being realized based on various materials systems as listed.

## 2. Intrinsic magnetic topological insulator

Intrinsic magnetic TIs provide an excellent platform for the study of exotic quantum states, such as QAH states, chiral Majorana fermions, axion states and so on [33-39], arising from the interplay between band topology and magnetism. Among them, QAH effect is of fundamental importance in the field of spintronics due to its non-dissipative properties in transport. One approach to realize it is to find a 2D TI that comprises long-range magnetic order. Introducing magnetism into the 2D TI can break the time-reversal symmetry, such that one direction of spin channels will be canceled. Although QAH effect has been proposed theoretically in the last century [41], it is until 2013 when quantized edge resistance ($h/e^2$) was experimentally observed on Cr-doped (Bi, Sb)$_2$Te$_3$ thin films [15]. The chemical doping results into inhomogeneity in the band structure (gap, carrier density) and consequently extremely low quantization temperature. Therefore, intrinsic magnetic states of matter with uniform long-range magnetic order are highly desired.

As first discussed in the theoretical proposal of antiferromagnetic TIs in 2010 [52], both time-reversal symmetry $\Theta$ and fractional translation $T_{1/2}$ are broken but the combination $S = \Theta T_{1/2}$ is preserved in AFM TI, leading to a topologically nontrivial phase which shares with 3D strong TI similar topological $Z_2$ invariant and quantized magnetoelectric effect. The material realization of an intrinsic AFM TI was not initiated until 2017. "Magnetic extension" picture proposed that by inserting MnTe bilayer into the quintuple layer of Bi$_2$Te$_3$, the system tends to form septuple layers of MnBi$_2$Te$_4$, hosting a robust QAH state [53,54]. The material was first experimentally realized by molecular beam epitaxy (MBE) [55]. Subsequent theoretical works revealed its colorful physics and properties [56-59]. Since the successful preparation of single crystal MnBi$_2$Te$_4$, the surge of intrinsic magnetic TIs based on MnBi$_2$Te$_4$ family started. Following the discovery of MnBi$_2$Te$_4$, a series of superlattices of this family were discovered, denoted as MnBi$_2$Te$_4$·(Bi$_2$Te$_3$)$_n$ (n=1,2,3) [60-62]. In addition, we will briefly introduce other intrinsic magnetic TI candidates such as MnSb$_2$Te$_4$·(Bi$_2$Te$_3$)$_n$ (n=1,2) and EuSn$_2$As$_2$ families.

### a) MnBi$_2$Te$_4$·(Bi$_2$Te$_3$)$_n$

In 2013, Lee *et al.* [63] synthesized the polycrystalline powder of MnBi$_2$Te$_4$ by the flux-method. In 2017, from MBE growth of heterostructure composed of MnSe and Bi$_2$Se$_3$, it was found that the topological surface state of this structure is located on the surface of the whole system, rather than at the interface of the two materials like other topological heterostructures. It was realized that the layered structure of MnSe and Bi$_2$Se$_3$ is a new type of single crystal, MnBi$_2$Se$_4$. Such transformation is also applicable to MnBi$_2$Te$_4$ [53,54,64,65], and it is MnBi$_2$Te$_4$ which is the focus of intrinsic magnetic TI study due to its desirable magnetic, electronic, and structural properties.

The structure of MnBi$_2$Te$_4$ was refined to be in the hexagonal space group $R\bar{3}m$ (No. 166) [60].

Its minimum structural unit is composed of seven atomic layers with stacking order Te-Bi-Te-Mn-Te-Bi-Te, which is called a septuple-layer (SL) and the adjacent layers are bonded by van der Waals force, as shown in Fig.2(a). The unit cell of $MnBi_2Te_4$ is composed of three SLs stacked in the -A-B-C- fashion, and its lattice constant $c$ is about 4.07 nm. Its Neel temperature $T_N(124) \approx 24.4\ K$ [66], above which the AFM order is transformed into paramagnetism (PM) (Fig. 2b). Neutron diffraction experiments point out that the ground state magnetic structure of $MnBi_2Te_4$ is the $A$-type AFM phase [66,67]. The magnetic moment of each SL points out of plane, and the magnetic moments of adjacent layers are opposite. Of course, if $Bi_2Te_3$ quintuple-layers (QLs) is inserted between SLs, we can get $MnBi_4Te_7$, $MnBi_6Te_{10}$, and $MnBi_8Te_{13}$ superlattices [60-62,68]. $MnBi_4Te_7$ can be regarded as a sandwich structure formed by inserting one QL into each SL. Similarly, $MnBi_6Te_{10}$ and $MnBi_8Te_{13}$ are formed by inserting two or three QLs in each SL respectively. Note that the space group of $MnBi_2Te_4$, $MnBi_6Te_{10}$, and $MnBi_8Te_{13}$ is $R\bar{3}m$, but the space group of $MnBi_4Te_7$ is $P\bar{3}m1$. Since the distance between two SLs in $MnBi_4Te_7$ and $MnBi_6Te_{10}$ is larger than that in $MnBi_2Te_4$, their interlayer AFM coupling is weaker. The results of magnetic transport measurement show that the AFM-PM transition temperature of $MnBi_4Te_7$ is $T_N(147) \approx 12\ K$ (Fig. 2c) and that of $MnBi_6Te_{10}$ is $T_N(1610) \approx 10.7\ K$ (Fig. 2d). More interestingly, with further increasing SL spacing, the compound of $MnBi_8Te_{13}$ has become the first intrinsic FM TI with $T_C(1813) \approx 10.5\ K$ (Fig. 2e). The lattice constants and magnetic transition temperatures of these different compounds are also summarized in Fig. 2f.

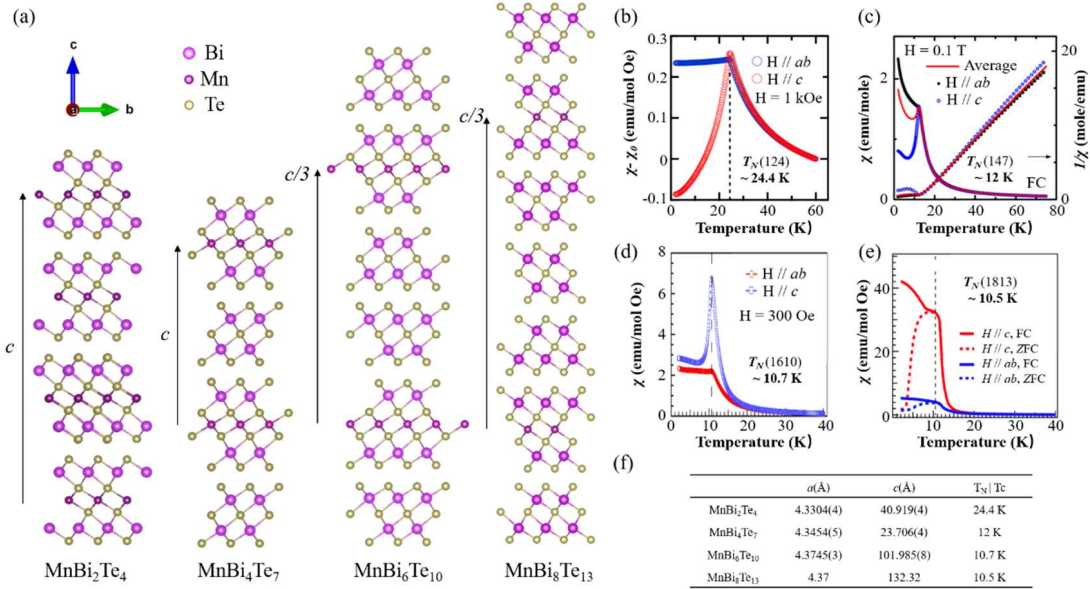

FIG. 2. (a) Crystal structure of $MnBi_2Te_4 \cdot (Bi_2Te_3)_n$. (b-e) Magnetic properties of $MnBi_2Te_4 \cdot (Bi_2Te_3)_n$ (n = 0,1,2,3) [69-72]. (f) Summary of lattice constants for $MnBi_2Te_4 \cdot (Bi_2Te_3)_n$.

The band structure of MnBi$_2$Te$_4$, as the first intrinsic magnetic TI, has been intensively studied [59,69,73-76] and the TSS inside the bulk gap is the focus of attention. At the early stage, a sizable gap was found for the TSS Dirac cone with temperature-independent behavior [59,77,78]. However, subsequent ARPES works with systematic photon-energy-dependent measurement and higher energy and momentum resolution have revealed the nearly gapless behavior of TSS [69,73-76,79-83], showing sample and location dependence (Fig. 3a). Here we use the term "nearly gapless" to describe the experimental observation that the size of Dirac gap varying from being vanishing to dozens of millielectronvolts, being much smaller than expected by theoretical calculation [56-59]. Such behaviors suggest much reduced effective magnetic moments felt by the TSS, which may arise from surface magnetic reconstruction or TSS redistribution (extension to the bulk). Currently there are several proposed mechanisms which may lead to one of these two phenomena yet none of them has been experimentally validated. Please refer to our recent review for more details [33].

Since the SLs and QLs in the heterostructure members of this family (MnBi$_4$Te$_7$, MnBi$_6$Te$_{10}$, MnBi$_8$Te$_{13}$) are combined by van der Waals forces, there are different terminations after cleaving the sample. As shown in Fig. 3b-d, the band structure on S-termination is very similar to that of MnBi$_2$Te$_4$, and the band structure on Q- and double Q-terminations show hybridization features between the TSS and certain bulk bands [70,71,79,84-92]. Again, no signature of sizable magnetic gap can be found for the TSS from all the different terminations of AFM members. The sizable magnetic gap of TSS was realized based on the S-termination of FM MnBi$_8$Te$_{13}$, with the gap size decreasing monotonically with increasing temperature and closing right at the Curie temperature [72].

Although the lack of sizable magnetic gap of TSS obscures the realization of topological quantized transport at high temperature (say, at the level of AFM transition temperature), QAH effect has indeed been realized at low temperature (1.4 $K$, Fig. 3e) based on 5 QLs films of MnBi$_2$Te$_4$, key evidence of a 2D Chern insulator [93]. The characteristics of an axion insulator state were also observed at zero magnetic field based on 6 SLs [45]. Under a perpendicular magnetic field (15 $T$), characteristics of high-Chern-number quantum Hall effect without Landau levels and contributed by dissipationless chiral edge states are observed, indicating a high Chern number Chern insulator with $C = 2$ (9, 10 SLs) [94]. The A-AFM configuration exhibits layer Hall effect in which electrons from the top and bottom layers deflect in opposite directions due to the layer-locked Berry curvature, resulting in the characteristic of the axion insulator state (6 SLs) [95]. We envision that half quantized Hall transport at the level of 10 $K$ can be realized based on the S-termination of FM MnBi$_8$Te$_{13}$ with sizable magnetic TSS gap [72].

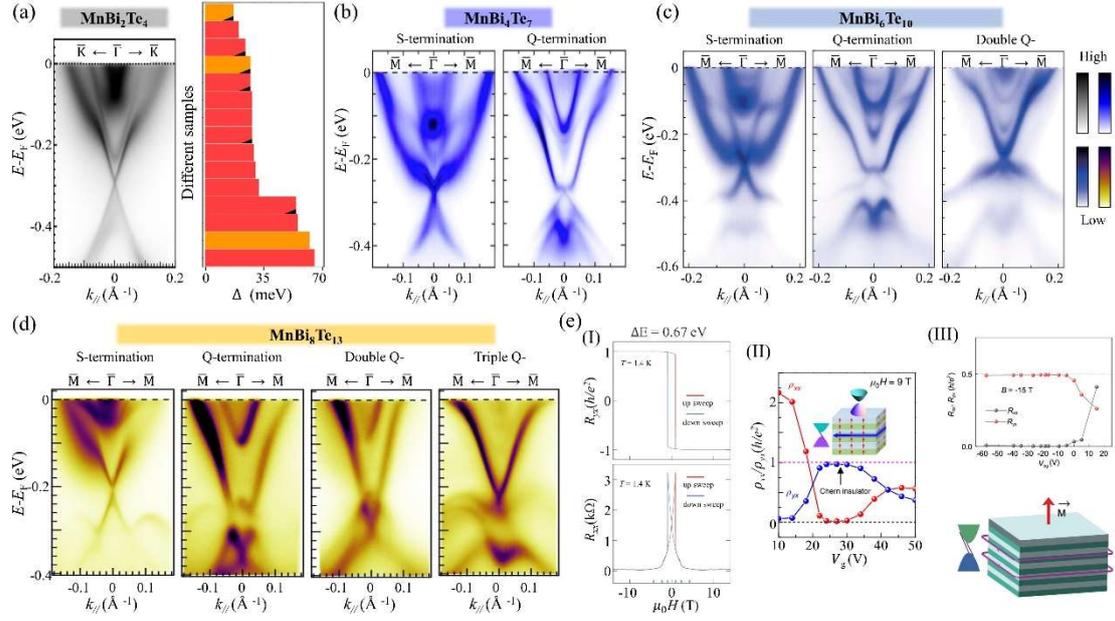

FIG. 3. (a) Observation of nearly gapless TSS in $MnBi_2Te_4$ single crystal (0001) surface (left, from [69]) and the variation of TSS gap in different samples (right, from [83]). (b, c, d) APRES spectra measured from the S-and Q- (double Q-, triple Q-) terminations of $MnBi_4Te_7$ [96], $MnBi_6Te_{10}$ [79] and $MnBi_8Te_{13}$ [72], respectively. (e) Observation of QAH effect (I, from [93]), axion insulator phase (II, from [45]) and high-Chern number Chern insulator (III, from [94]) based on $MnBi_2Te_4$ films with different number of layers.

### b) MnSb$_2$Te$_4$·(Sb$_2$Te$_3$)$_n$

Since the successful synthesis of MnBi$_2$Te$_4$·(Bi$_2$Te$_3$)$_n$ single crystals, elemental substitutions have been explored in order to manipulate its magnetic and electronic properties. It turns out the Bi site can be completely substituted by Sb atoms. The resulting MnSb$_2$Te$_4$·(Sb$_2$Te$_3$)$_n$ family of materials are currently under intensive investigation. Theoretically, this family ($n = 0, 1, 2$) is also predicted to host similar AFM ground state and AFM TI phase [97,98], yet there lacks consistency between/among experiments and calculations on the exact magnetic ground state and band topology of MnSb$_2$Te$_4$ [99-105]. Notably, ARPES results reveal significant hole doping for all the members studied so far, leaving the detailed TSS Dirac cone structure not straightforward to study [104,106]. The crystal structure of MnSb$_4$Te$_7$ adapts a space group of $P\bar{3}m1$. The Mn layer constitutes a long-range magnetic order with moments along the $c$ direction (Fig. 4a,b) [106] ($A$-type AFM with $T_N = 13.5\ K$). ARPES measurement also reveals hole doping for the band structure with expected Dirac cone located at $180\ meV$ above the Fermi level (Fig. 4c). Pressure experiments and DFT calculations have revealed multiple topological phases corresponding to various magnetic structures and the emergence of superconductivity (Fig. 4d) [97,105-109]. Similar hole doping and multiple magnetic topological phases have also been found in MnSb$_6$Te$_{10}$, a ferromagnetic member of this family at its ground state (Fig. 4f,g) [110]. Considering the universal electron doping behavior in MnBi$_2$Te$_4$·(Bi$_2$Te$_3$)$_n$ family, it is natural to expect carrier tunability and magnetic manipulation based on the mutual substitution of Sb and Bi in Mn(Bi, Sb)$_2$Te$_4$·((Bi, Sb)$_2$Te$_3$)$_n$ series. In fact, a tunable TSS Dirac gap varying from being gapless to larger than $100\ meV$ has been reported in Sb doped MnBi$_2$Te$_4$, with its gap size proportional to the doping level [111].

Except MnBi$_2$Te$_4$·(Bi$_2$Te$_3$)$_n$ and MnSb$_2$Te$_4$·(Sb$_2$Te$_3$)$_n$ families, it is noted that MnBi$_2$Se$_4$ in the $R\bar{3}m$ space group shares the same magnetic and topological properties of MnBi$_2$Te$_4$. This phase turns out to be unstable in the bulk crystal form. Recent efforts have succeeded in synthesizing ultrathin films of MnBi$_2$Se$_4$ using nonequilibrium molecular beam epitaxy [112]. Its magnetic structure, however, deviates from the expected A-AFMz structure and the response of TSS Dirac cone to the magnetic order remains to be investigated.

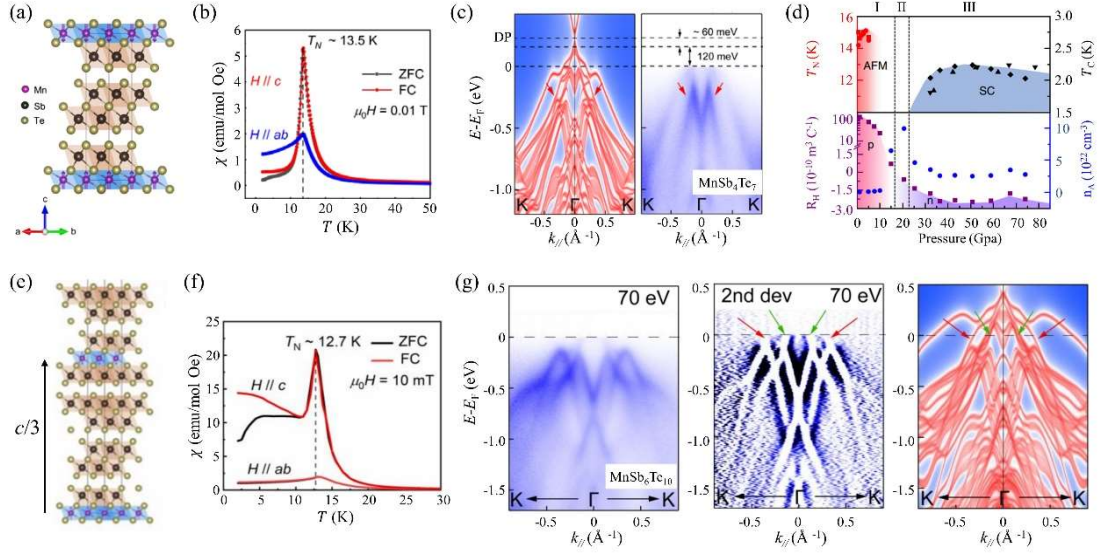

FIG. 4. (a), Crystal structure (a, c), magnetic transport properties (b, f) and band structure (c, g) of MnSb$_4$Te$_7$ and MnSb$_6$Te$_{10}$, respectively [106,110]. (d) The pressure dependence of superconducting transition temperature $T_C$, AFM transition temperature $T_N$ (upper panel), Hall coefficient $R_H$ and carrier concentration (lower panel) at 10 K (different symbols represent different samples in the upper panel) [109].

c) EuM$_2$X$_2$ (M = metal; and X = Group 14 or 15 element)

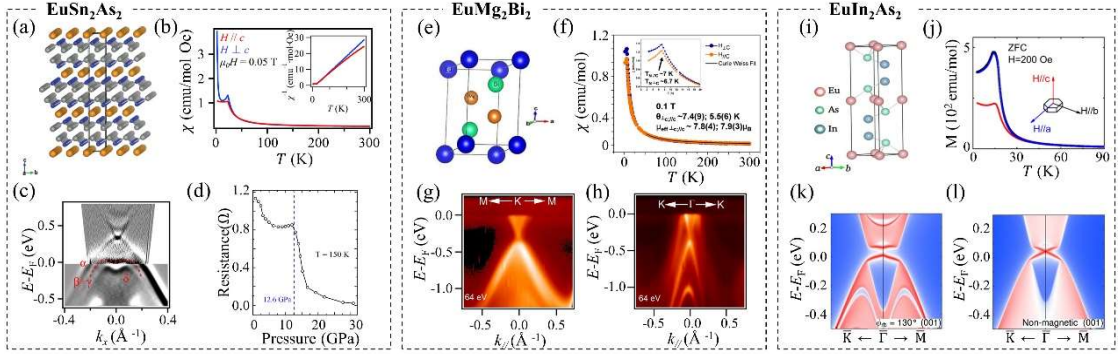

FIG. 5. (a, e, i) Crystal structures of EuSn$_2$As$_2$ [113], EuMg$_2$Bi$_2$ [114] and EuIn$_2$As$_2$ [115], respectively. In (a), Eu atoms are shown in orange, Sn in gray and As in bule. (b, f, j) Magnetic transport properties of EuSn$_2$As$_2$ [74], EuMg$_2$Bi$_2$ [116] and EuIn$_2$As$_2$ [117], respectively. (c) Band structure of EuSn$_2$As$_2$ measured by pump-probe ARPES [74]. (d) The pressure dependence of resistance for EuSn$_2$As$_2$ [118]. (g, h) Band structure of EuMg$_2$Bi$_2$ along M-K-M and K-Γ-K, respectively [114]. (k, l) Calculated band structure of EuIn$_2$As$_2$ [119].

EuSn$_2$As$_2$ belongs to the group of compounds with formula AM$_2$X$_2$ (A = alkali, alkaline earth, or rare earth cation; M = metal; and X = Group 14 or 15 element). Here we focused on the A = Eu compounds with intrinsic AFM order. The M site can be occupied by various types of metals such as Mg, In and Sn. EuSn$_2$As$_2$, as an important member in intrinsic magnetic TI family, crystallizes in the hexagonal space group $R\bar{3}m$. The Eu atoms are triangularly distributed and sandwiched by two honeycomb SnAs layers to form a layered structure (Fig. 5a). The magnetic moment provided by Eu atom forms an *A*-type AFM configuration with $T_N = 25\ K$ [74,113] (Fig. 5b). ARPES measurements have revealed a TSS Dirac cone locating ~0.4 $eV$ above the Fermi level at the PM phase, suggesting a strong 3D TI phase (Fig. 5c). Yet no observable change of the TSS or carrier concentration can be found in the AFM state, indicating weak coupling between the Eu moments and low-energy bands [74,120]. Magnetic property and transport measurements report negative magnetoresistance and complicated magnetic transitions from an AFM state to a canted ferromagnetic (FM) state and then to a polarized FM state as the magnetic field increases [120,121]. Electrical resistance measurements under pressure reveal an insulator-metal-superconductor transition at low temperature around 5 and 15 $GPa$ (Fig. 5d). A new $C2/m$ phase appears when the pressure is higher than 14 $GPa$. As the pressure continues to increase, the superconductivity persists up to 30.8 GPa with $T_C$ maintaining a constant value ~ 4 $K$ [118]. It is also found that the pressure has an enhancement effect on the AFM transition temperature and negative magnetoresistance [122].

For EuMg$_2$Bi$_2$, it crystallizes into the tetragonal CaAl$_2$Si$_2$ structure type with space group $P\bar{3}m1$ (No. 164) [116] (Fig. 5e). Magnetic property measurements revealed AFM transition temperature $T_N \sim 7\,K$ with slight anisotropy and positive Curie-Weiss temperature indicating ferromagnetic interaction between Eu atoms ($7.8\,\mu_B$) (Fig. 5f). Like Mn-Bi-Te family, AFM configuration between FM layers of Eu is established. The difference is that the moments point out-of-plane in Mn-Bi-Te but in-plane for EuMg$_2$Bi$_2$. ARPES measurements and DFT calculations have revealed Dirac surface state features and nontrivial band topology (Fig. 5g-h), suggesting EuMg$_2$Bi$_2$ as a magnetic topological insulator candidate [114,116].

EuIn$_2$As$_2$ crystallizes into the hexagonal space group $P6_3/mmc$, containing layers of Eu$^{2+}$ cations separated by In$_2$As$_2^{2-}$ layers along the crystallographic $c$-axis [123] (Fig. 5i). Magnetic property and neutron diffraction measurements have determined a colinear AFM ground state with the moments lying in the $ab$-plane [117,119,123,124] (Fig. 5j). Furthermore, a complicated broken helix order is reported by neutron diffraction, tripling the unit cell along $c$-axis. EuIn$_2$As$_2$ was predicted as a high-order topological axion insulator candidate [119,125] protected by the magnetic crystalline symmetry. Such a state has gapless TSS Dirac cone at the symmetry-protected termination and gapped ones at other surfaces (Fig. 5k-l). However, like other $AM_2X_2$ compounds, its hole-doping nature as observed by ARPES [115,117] and STM [126] has prevented the detailed study on the TSS band structure, especially the gap behavior. Further chemical and band structure engineering are strongly called for to tune the chemical potential for access to the TSS Dirac point in this family.

There are also theoretical calculations which predict materials such as several Eu$_5M_2X_6$ ($M$ = metal, $X$ = pnictide) Zintl compounds [127,128], 2D EuCd$_2$Bi$_2$ [129], NiTl$_2$S$_4$ [130] and so on to be intrinsic magnetic TI candidates yet their growth, band structure, magnetic structure and band topology remain to be investigated.

## 3. Magnetic Weyl/Dirac semimetals

In a Dirac semimetal, two doubly degenerate bands contact at discrete momentum points called Dirac points and disperse linearly along all directions around these points. The four-fold degenerate Dirac points need symmetries to ensure their existence, such as time-reversal symmetry $T$, inversion symmetry $P$, rotational symmetry and nonsymmorphic symmetry. In a Dirac semimetal with $TP$ symmetry, when either $T$ or $P$ is broken, each doubly degenerate band is lifted, so that the Dirac cones can split into multiple Weyl cones, giving birth to Weyl semimetals. However, in 3D systems with AFM order that breaks both $T$ and $P$ but respect their combination $PT$, four-fold degenerate Dirac points can still exist, resulting into AFM DSM [131]. Such consideration has also been generalized to 2D systems [132-134].

In magnetic Weyl semimetals, spin-polarized conduction and valence bands touch at finite number of nodes, forming pairs of Weyl nodes. In each pair, the quasiparticles carry opposite chirality and can be viewed as the "source" ("+" chirality) and the "sink" ("-" chirality) of the Berry curvature. Odd pairs of Weyl nodes with opposite chirality can be expected in systems with $T$ symmetry breaking, such as $Co_3Sn_2S_2$ [135,136] and $Mn_3X$ (X=Sn, Ge) [137-139]; While for systems with time-reversal symmetry $T$, the total number of Weyl nodes pairs must be even. Noncentrosymmetric WSMs belong to this category, such as TaAs family [140-143]. If $P$ and $T$ symmetries are both preserved, Weyl nodes with opposite chirality can merge at the same momentum and form a four-fold Dirac point (assisted by additional crystal symmetry), such as $Na_3Bi$ [21,144] and $Cd_3As_2$ [23,145,146]. Due to non-zero Berry curvature, many novel physical properties such as giant anomalous Hall effect and giant anomalous Hall angle, chiral anomaly, anomalous Nernst effect will emerge in magnetic Weyl semimetal, holding potential applications in spintronics field. In the early stage, several candidate materials were predicted, such as $R_2Ir_2O_7$ (R=Nd,Pr) [20], $HgCr_2Se_4$ [147]. Recent efforts have focused on $Co_3Sn_2S_2$ [135,136] and $Mn_3X$ (X=Sn, Ge) [137-139] which clearly host the band structure and transport characters as expected by magnetic WSM. We will briefly introduce these magnetic materials.

### a)  FeSn

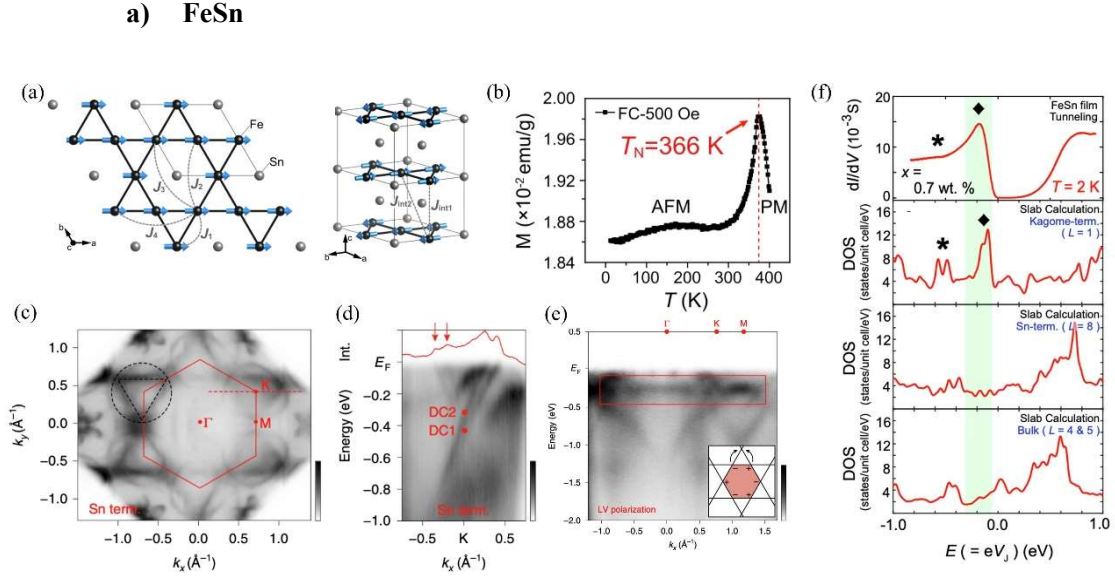

FIG. 6. (a) FeSn lattice structure and magnetic configuration, from [148]. (b) Magnetization as a function of temperature under field cooling (FC) with an applied magnetic field of 500 Oe, from [149]. (c, d, e) ARPES Fermi surface mapping and spectra reveal two Dirac cones features around $K$ point (d) and flat band close to the Fermi level (e), from [150]. (f) Planar tunneling spectroscopy reveals an anomalous enhancement in tunneling conductance within a finite energy range of FeSn (black diamond), attributed to a spin-polarized flat band, from [151].

FeSn crystallizes in a hexagonal structure ($P6/mmm$) with the Fe atoms forming a Kagome lattice [149,150,152,153]. Like $Fe_3Sn_2$, FeSn is formed by interlacing $Fe_3Sn$ layer and Sn layer. The difference is that there is only one Kagome layer ($Fe_3Sn$ layer) in a unit cell (Fig. 6a). It is closer to the two-dimensional limit than $Fe_3Sn_2$. Below $T_N = 365\ K$ (Fig. 6b), the Fe spins form ferromagnetic Kagome layers which are stacked antiferromagnetically along the $c$ axis. The Dirac nodal line along the K-H line opens small energy gaps when SOC is considered, except at the H point where a gapless Dirac point (protected by $PT$ and $S_{2z}$ symmetry) still exist, rendering FeSn as an AFM DSM. Such gapless Dirac cones have been directly observed by ARPES [149,150] (Fig. 6c-d). Besides, the flat band because of the Kagome layer has also been observed directly by ARPES (Fig. 6e). Furthermore, in a planar tunneling spectroscopy measurement [151], an anomalous enhancement in tunneling conductance within a finite energy range of FeSn has been observed in its Schottky heterointerface with Nb-doped $SrTiO_3$ (Fig. 6f). Such tunneling conductance peak is attributed to spin-polarized flat band localized at the ferromagnetic Kagome layer at the Schottky interface.

### b) Co$_3$Sn$_2$S$_2$

Co$_3$Sn$_2$S$_2$ crystallizes in the $R\overline{3}m$ space group with a stacking order -Sn-S-Co$_3$Sn-S- from top to bottom. The central Co layer forms a two-dimensional Kagome lattice with one Sn atom at the center of the hexagon, as shown in Fig. 7a. Co$_3$Sn$_2$S$_2$ is a ferromagnet with a curie temperature of $175\ K$ and a magnetic moment of $0.3\ \mu_B/Co$. In magnetization measurement, the saturation field along $c$ axis is low ($\sim 0.05\ T$) but along in-plane is extremely high ($> 9\ T$), confirming that the easy magnetic axis is $c$-axis [48,154]. Combining theory and experiments, Co$_3$Sn$_2$S$_2$ is an ideal ferromagnetic Weyl semimetal with three pairs of Weyl points whose energies are only $\sim 60\ meV$ above the Fermi level [135,136,155-160]. The Weyl nodes have been observed by ARPES after doping alkaline metal (Fig. 7c). Three Fermi arcs form a triangular-like loop around the K' point near Fermi surface. Meanwhile, the electronic structure doesn't undergo obvious dispersion along the $k_z$ direction, suggesting the nature of topological surface states (Fig. 7b). Termination-dependent surface band structures of Co$_3$Sn$_2$S$_2$ were observed by using STM [135]. Different surface potentials imposed by three different terminals will change the Fermi arc contour and Weyl node connectivity. On the Sn-termination, the Fermi arcs connect Weyl nodes within the same Brillouin zone, while on the Co-termination, the connectivity spans the two adjacent Brillouin zones. On S-termination, Fermi arcs overlap with the trivial surface-projected bulk bands. The topologically protected and unprotected electronic properties of Weyl semimetals Co$_3$Sn$_2$S$_2$ were verified.

According to first-principles calculation, the Weyl nodes in Co$_3$Sn$_2$S$_2$ locate close to the Fermi level and produce a giant AHC ($\sim 1100\ \Omega^{-1}cm^{-1}$), which has been directly observed in transport measurement (Fig. 7d-e) [48,160,161]. Besides, giant anomalous Hall angle also emerges in this material. As shown in Fig. 7e, with increasing temperature, a maximum value of nearly 20% is reached around $120\ K$, which is at least one order of magnitude higher than that of conventional magnetic materials. Negative magnetoresistance is found in Co$_3$Sn$_2$S$_2$, as shown in Fig. 7f, when the magnetic field is applied in the in-plane direction, the longitudinal resistance is negative, and when the external magnetic field is applied in the out-of-plane direction, the longitudinal resistance changes from negative to positive, showing evidence of chiral anomaly [48,154,160,161]. In Co$_3$Sn$_2$S$_2$ thin film, a maximum Nernst thermopower of $\sim 3\ \mu V K^{-1}$ is achieved [50], demonstrating the possibility of application of hard magnetic topological semimetals for low-power thermoelectric devices.

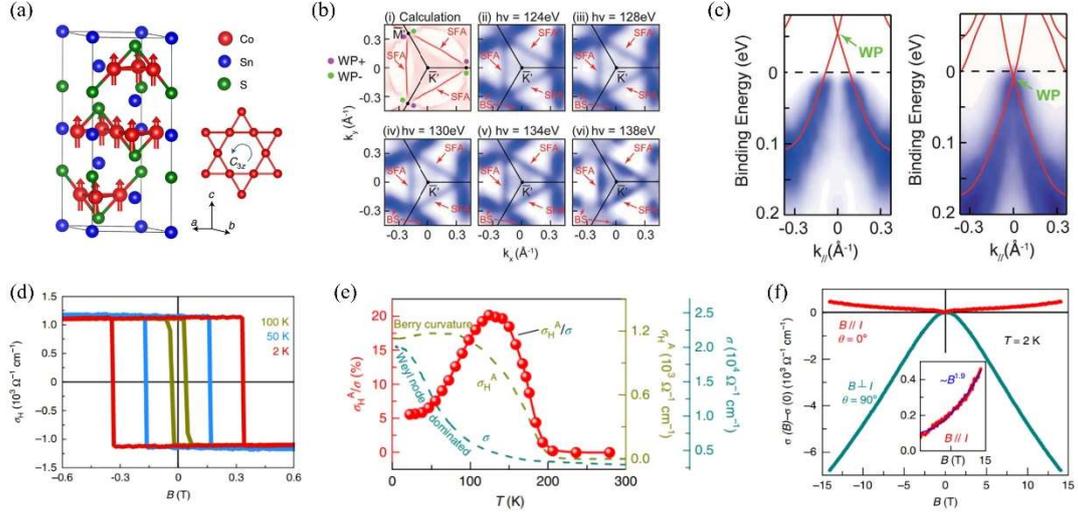

FIG. 7. (a) $Co_3Sn_2S_2$ lattice structure and magnetic configuration, from [48]. (b) Calculated Fermi surface (i) and experimental Fermi surfaces under different photon energies (ii-vi) around K' points in $Co_3Sn_2S_2$, SFA: surface Fermi arc, BS: bulk state, from [136]. (c) Intrinsic band structure (left), and band structure after potassium dosing and from calculation (red lines) of $Co_3Sn_2S_2$, from [136]. (d) Field dependence of the Hall conductivity $\sigma_H$, from [48]. (e) Temperature dependences of the anomalous Hall conductivity ($\sigma_H^A$), the charge conductivity ($\sigma$) and the anomalous Hall angle ($\sigma_H^A/\sigma$) at zero magnetic field, from [48]. (f) Measured magnetoconductance for B⊥I and B//I, from [48].

### c) Mn$_3$X (X = Sn, Ge)

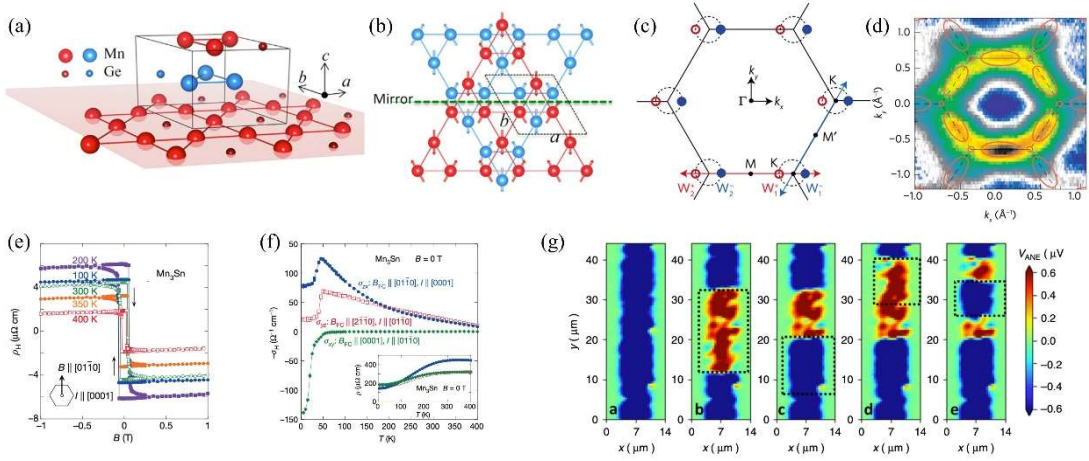

FIG. 8. (a) Mn$_3$Ge lattice structure and magnetic configuration (b), from [138]. (c) Band structure calculation reveals the existence of one pair of Weyl points close to the Fermi level around K points in Mn$_3$Sn, from [137]. (d) Fermi surface mapped by ARPES and from calculation (purple curves) of Mn$_3$Sn, from [137]. (e, f) Field dependent Hall resistance at different temperatures show AHE behavior and temperature dependent zero-field component of the AHE in Mn$_3$Sn, from [139]. (g) Anomalous Nernst voltage $V_{ANE}$ image mapped by scanning thermal gradient microscopy reveal the existence of magnetic domain and domain writing, from [162].

Mn$_3$X (X = Sn, Ge) has a hexagonal Ni$_3$Sn-type structure and crystalizes in the $P6_3/mmc$ space group. One unit cell consists of two sets of Mn layers stacked along $c$-axis and each Mn layer forms a breathing-type Kagome lattice with one Sn atom at the center of the hexagon, as shown in Fig. 8a. Mn$_3$Sn and Mn$_3$Ge are both chiral antiferromagnets which means Mn moments are forming a 120° ordering with a negative vector chirality (Fig. 8b) [163,164]. The AFM transition temperature of Mn$_3$Sn and Mn$_3$Ge is 430 $K$ and 372 $K$, respectively. Because the electronic structures of Mn$_3$Sn and Mn$_3$Ge are quite similar and the study of Mn$_3$Sn is more comprehensive, we will mainly focus on Mn$_3$Sn. Mn$_3$Sn possesses non-collinear AFM spin texture and strong SOC effect, which produce multiple pairs of Weyl points close to the Fermi level, according to first-principal calculation [137,165,166] (Fig. 8c). However, ARPES spectra [137] measured on Mn$_3$Sn lacks clear features of quasiparticle bands, likely due to strong correlation effect of Mn $3d$ electrons (Fig. 8d).

Novel transport properties governed by the topological nature can serve as evidence for Weyl fermions. In Mn$_3$Sn, strongly anisotropic magnetoconductance was observed. The sign of magnetoconductance changed when rotated the direction of magnetic field from being parallel to perpendicular to the current direction, serving as strong evidence of chiral anomaly [137,163]. The large AHE is also a key characteristic of magnetic WSM. In the traditional sense, because the magnetic configuration of Mn$_3$Sn is AFM, there is no net magnetic moment in this material and the AHE will not emerge. But many reports revealed that Mn$_3$Sn exhibits a large AHE [138,139,165,167-170]. Fig. 8e-f shows the temperature-dependence of zero-field Hall conductivity under different magnetic field and current directions [139]. We can see that when the magnetic field and the current are applied along the $(01\bar{1}0)$ and $(0001)$ direction, the $\sigma_H$ will achieve a maximum value of nearly $130\ \Omega^{-1}cm^{-1}$ at $50\ K$. In Mn$_3$Ge, by employing similar magnetic field and current direction, even higher AHC have been obtained [138,167]. The large AHE in Mn$_3$X is mainly caused by the non-zero Berry curvature produced by Weyl nodes [163]. Besides chiral anomaly and large AHE, many other exotic physical properties such as large anomalous Nernst effect [162,163,171-173], planer Hall effect [170,174,175], magnetic spin Hall effect and magnetic inverse spin Hall effect [176] are also observed in Mn$_3$X. Furthermore, as shown in Fig. 8g, anomalous Nernst voltage $V_{ANE}$ image mapped by scanning thermal gradient microscopy reveals the existence of magnetic domains. The orientation of these domains can be changed (written) by laser-induced local thermal gradient [162], offering a chance to study spintronics phenomena in non-collinear antiferromagnets with spatial resolution.

### d) Co$_2$MnGa

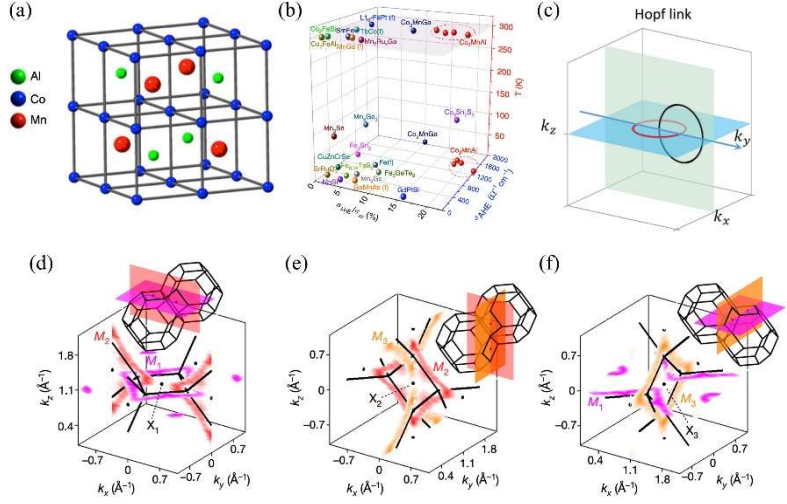

FIG. 9. (a) Crystal structure of Co$_2$MnAl [177]. (b) Comparison of anomalous Hall angle tan$\Theta^H$ = $\sigma_{AHE}/\sigma_{xx}$ and anomalous Hall conductivity $\sigma_{AHE}$ between Co$_2$MnAl and other magnetic conductors [177]. (c) Hopf link which consists of two rings on the mirror planes and intertwined each other [178]. (d)-(f) Linked Weyl loops in Co$_2$MnGa [179]. M1-and M2-loop Fermi surfaces plotted in adjacent bulk Brillouin zones (d). Same as (d) but for the M2-and M3-Fermi surfaces (e) and the M1-and M3-Fermi surfaces (f).

A new family of magnetic WSM emerged among the magnetic Heusler alloys, i.e., the Heusler alloy WSMs [180,181]. It's an important family due to their rich transport properties and several superiorities. Firstly, the Curie temperatures of most Heusler compounds are above the room temperature [182,183]. Secondly, this kind of materials has a significant anomalous Hall effect and spin Hall effect arising from the large Berry curvature [178,180,181,184-186]. Thirdly, Heusler compounds are usually soft magnetic materials, which means that their magnetization direction can be tuned by a weak magnetic field. These properties facilitate spin manipulation and applications in spintronics, as a result, these Heusler alloy WSMs have been widely studied.

As full Heusler compounds, Co-based Heusler materials have the formula of Co$_2$XZ (X = IVB or VB; Z = IVA or IIIA), here we focus on Co$_2$MnGa and Co$_2$MnAl. Co$_2$MnGa (Co$_2$MnAl) crystallizes in a face-centered cubic Bravais lattice (space group $Fm\bar{3}m$, No. 225), as shown in Fig. 9a. The relevant symmetries are the three mirror planes and three $C_4$ rotation axes. The Curie temperature of Co$_2$MnGa and Co$_2$MnAl are known to be ~700 K [178] and 726 K [182], respectively. Transport experiments showed that Co$_2$MnAl has a giant room-temperature anomalous Hall effect with the Hall angle ($\Theta^H$) reaching a record value tan$\Theta^H$= 0.21 at the room temperature among magnetic conductors [177], as shown in Fig. 9b. This property results from the gapped nodal rings that generate large Berry curvature. Furthermore, for $Co_2MnGa$ films, when the $E_F$ is set in the magnetization-induced gap of the Weyl cones by the electronic doping, the highest anomalous Nernst thermopower of a record value 6.2 $\mu VK^{-1}$ will be reached at room temperature [187].

The Hopf link is originally a mathematical concept which consists of two rings on the two perpendicular planes, each passing through the center of each other, as shown in Fig. 9c. The symmetry of $Co_2MnGa$ can protect this band crossing associated with the unusual linking-number (knot theory) invariant, giving rise to a variety of new types of topological semimetals [178,179,184-186,188-190]. Systematic ARPES investigation of the electronic structure of $Co_2MnGa$ has been carried out and directly revealed three intertwined degeneracy loops in the material's three-torus bulk Brillouin zone (Fig. 9d-f). In addition, the Seifert boundary states protected by the bulk-linked loops have been predicted and observed, while the Links and knots in the electronic structure and the accompanied exotic behaviors remain unexplored.

### e) EuB$_6$

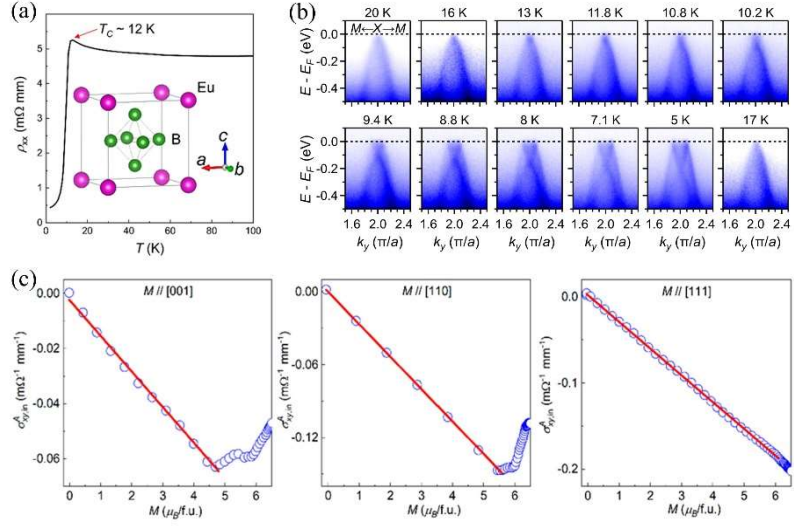

FIG. 10. Crystal structure of EuB$_6$ and its longitudinal resistivity as a function of temperature [191]. (b) Temperature dependent band structure of B-terminated surface along M-X, which is taken with $hv = 135\ eV$ [192]. (c) The intrinsic anomalous Hall conductivity as a function of different magnetization at $2\ K$ [191].

The EuB$_6$ crystallizes in a similar body-centered-cubic-like crystal structure with space group $Pm\bar{3}m$ (No. 221) (Fig. 10a). EuB$_6$ is a soft ferromagnetic semimetal which has a very small magnetic anisotropy energy so that the magnetization can be easily modulated by magnetic field [191,193-195]. Electronic transport and magnetic susceptibility measurements showed that the system undergoes a paramagnetic to ferromagnetic phase transition at about 15.3 $K$ and a new ferromagnetic phase manifests below about 12.5 $K$ with moment oriented to the (111) direction [196-198]. The magnetotransport properties of EuB$_6$ have been widely studied around magnetic phase transition point, such as the metal-insulator transition, colossal magnetoresistance and quantum nematic phase [199-201].

It has been predicted that EuB$_6$ is a topological nodal-line semimetal when the magnetic moment is aligned along the (001) direction, and it turns out to be a WSM with three pairs of Weyl nodes when rotating the magnetic moment to (111) direction. Specifically, when the moment is in the (110) direction, a composite semimetal phase featuring the coexistence of a nodal line and Weyl points manifests [193]. The electronic structures on the two different cleavage planes in EuB$_6$, i.e., the Eu- and B-terminated surfaces, have been investigated [192,202]. For the B-termination, in the FM state, obvious Zeeman splitting occurs for both the conduction and valence bands, which gives rise to the overlap of subbands and thus the band inversion at the time-reversal point X of the Brillouin zone (Fig. 10b). In this case, EuB$_6$ enters a topological semimetal state with an ideal electronic structure near $E_F$. The topological properties can be investigated by measuring the magnetotransport properties due to the correlation between the band structure and the local moments. Fig. 10c shows the intrinsic anomalous Hall conductivity as a function of magnetization with different directions at 2 $K$ [191]. An intrinsic large anisotropic magnetoresistance of -18% at 0.2 T was observed and interpreted as the modification from the Berry curvature in a tilted Weyl cone [203]. The theoretical prediction that a large-Chern-number quantum anomalous Hall effect could be realized in its (111)-oriented quantum-well structure [193] needs further investigations.

**f) Fe$_3$GeTe$_2$**

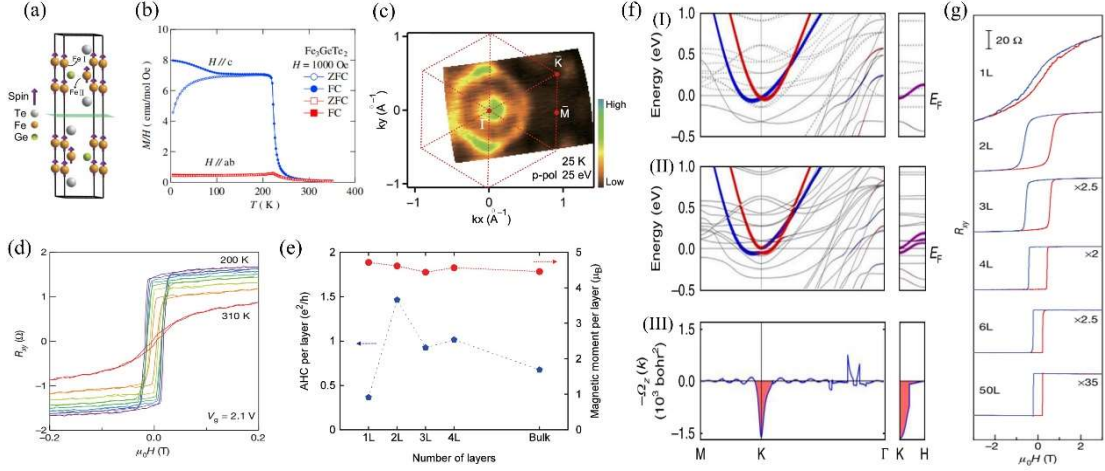

FIG. 11. (a) Fe$_3$GeTe$_2$ lattice structure, magnetic configuration and (b) magnetic properties, from [204,205]. (c) ARPES measured Fermi surface of Fe$_3$GeTe$_2$, from [205]. (d) Hall resistance of a four-layers Fe$_3$GeTe$_2$ flake [206]. (e) The dependence of AHC and magnetic moment per layer on the number of layers, from [207]. (f) Calculated electronic structures of Fe$_3$GeTe$_2$ without (I) and with (II) SOC. Majority spins: solid. Minority spins: dashed. (III) Calculated Berry curvature along the symmetry lines, from [208]. (g) Hall resistance with varying numbers of layers [206].

Fe$_3$GeTe$_2$ crystallizes in a hexagonal structure ($P6_3/mmc$, No. 194) in which the layered Fe$_3$Ge substructure are sandwiched by two layers of Te atoms (Fig. 11a). Fe$_3$GeTe$_2$ is ferromagnetic with Fe moments along the $c$ axis and a Curie temperature of $204 \sim 230\,K$ (Fig. 11b) [204,209-211]. ARPES measurements have revealed two pockets around Γ point and one at $K$ point (Fig. 11c). Temperature-dependent ARPES spectra exhibits a massive spectral weight transfer in the ferromagnetic state induced by exchange splitting [205]. Orbital-driven nodal line along K-H protected by crystalline symmetry has been predicted (Fig. 11f). Introducing SOC will gap the nodal line and generate large Berry curvature ([208]), an effective source of a large AHE in Fe$_3$GeTe$_2$. We note that Fe$_3$GeTe$_2$ is considered as a gapped nodal line semimetal with the Weyl point awaiting verification.

Fe$_3$GeTe$_2$ also contains very rich physical properties. Due to the gapped nodal line, negative magnetoresistance [212,213], anomalous Nernst effect [214] and anomalous Hall effect were observed [207,208,211]. Compared with other itinerant ferromagnetic materials, Fe$_3$GeTe$_2$ has both large anomalous Hall factor and anomalous Hall angle (Fig. 11d-e). Due to the weak interlayer coupling, Fe$_3$GeTe$_2$ can be exfoliated into sheets with different number of layers. More importantly, its novel transport and magnetic properties show stability at room temperature and dependence on the number of layers, interlayer coupling and carrier density [206,207,215-221], holding potential in spintronics applications.

## g) EuCd$_2$As$_2$

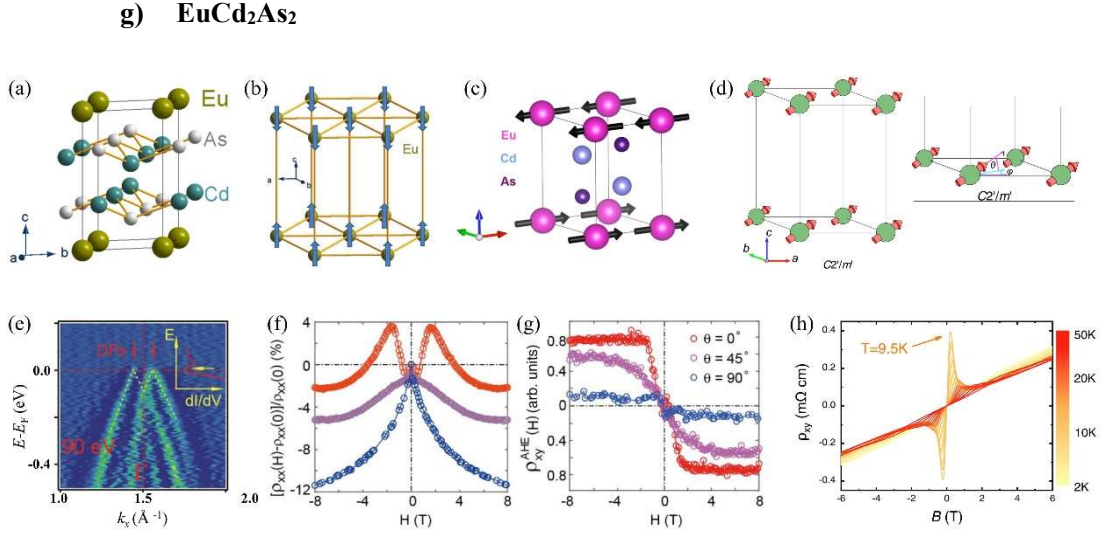

FIG. 12. (a) Crystal structure of EuCd$_2$As$_2$, from [222]. (b) Proposed A-type antiferromagnetic structure on Eu sites with the moments lying out-of-plane, from [222]. (c) Proposed A-type antiferromagnetic structure with the moments lying in-plane, from [223]. (d) Best-fit magnetic structure from neutron diffraction measurement with moments along the (210) direction with 30° canting, from [224]. (e) ARPES spectral along $Z - \Gamma - Z$ presents a "M"-shaped feature around Γ, from [225]. (f, g), Negative magnetoresistance and anomalous Hall resistance with three different orientations between $H$ and $E$, from [224]. (h) Magnetic-field dependence of the Hall resistivity at different temperatures shows giant nonlinear behavior, from [226].

EuCd$_2$As$_2$ belongs to EuM$_2$X$_2$ (M = metal; and X = Group 14 or 15 element) family in which several members are studied as magnetic TI candidates (see Section II). The exact band structure details and topological phase are sensitively related to the magnetic configuration. The crystal structure of EuCd$_2$As$_2$, with space group 164 ($P\bar{3}m1$), is shown in Fig. 12a. The Eu atoms form a simple hexagonal lattice at the 1a Wyckoff position. The As and Cd atoms at the 2b positions form the other four atomic layers with the sequence of -Cd-As-Eu-As-Cd- along the c axis [222,227,228]. Eu moments prefer an intralayer FM coupling and an interlayer AFM coupling along the $c$ axis, i.e., an A-type AFM (A-AFM), which doubles the unit cell along the $c$ direction. Fig. 12b and 12c show two such magnetic configurations by showing Eu atoms with magnetic moment directions along $c$ (A-AFMc) and along $a$ (A-AFMa). A-AFMc is proposed based on the anisotropic magnetic and transport properties [222,227]. A-AFMa is proposed based on the resonant elastic x-ray scattering [223,229], first-principles calculations [230] and magnetostriction measurements [231]. Furthermore, neutron diffraction on isotopic $^{153}$Eu and $^{116}$Cd revealed a $\mathbf{k} = (0, 0, 0)$ FM order at zero field with the Eu moments pointing along the in-plane (210) direction with a $\sim$ 30° out-of-plane canting (MSG $C2'/m'$, Fig. 12d) [224].

According to the first-principles calculation and symmetry analysis, various topological phases emerge based on different magnetic configurations in EuCd$_2$As$_2$. For A-AFMz, DSM phase exists with the gapless Dirac point protected by the $PTL$ symmetry operation which is the product of inversion symmetry $P$, time reversal symmetry $T$ and crystalline translation symmetry $L$ [225,232]. For A-AFMx, spin configuration breaks the $C_3$ symmetry in the AFM state of EuCd$_2$As$_2$ and leads to an axion insulator with a hybridization gap of $\sim 1\ meV$. Massless Dirac surface states appear on some surfaces protected by the mirror or $TL$ symmetries. For other surfaces without such symmetry, the surface states are gapped and the hinge states, associated with higher order TI states, emerge at the edges [125,233]. There are other calculations which predict EuCd$_2$As$_2$ as a WSM with a single pair of Weyl points very close to the Fermi level [224,228,234]. Such Weyl phase can be generated in EuCd$_2$As$_2$ by applying a magnetic field $> 1.5\ T$ along the $c$ axis [234] or alloying with Ba at the Eu site to stabilize the FM configuration [228]. In fact, the recently confirmed spin-canted structure as shown in Fig. 12d naturally hosts such Weyl semimetal phase [224]. Spectroscopically, ARPES measurements have observed linear band crossings at the Fermi level and especially an "M"-shaped feature around Γ point (Fig. 12e), suggesting a nontrivial band inversion. Such features cannot distinguish between the semimetal and insulator phase as the gap is only $\sim 1\ meV$, comparable to the thermal broadening effect at $\sim 3\ K$. ARPES or STS measurements at ultralow temperature are needed. Spin-resolved ARPES is also useful to examine the spin degeneracy of these linear bands and crossings.

Magnetic transport experiments have provided more information on the interplay between magnetism and band topology in EuCd$_2$As$_2$. Negative magnetoresistance (Fig. 12f), as signature of chiral anomaly is observed along with anomalous Hall effect (Fig. 12g) [223,224,226]. These transport results support as-grown EuCd$_2$As$_2$ in a semimetal phase, yet gate tunable transport is needed to verify the absence of gap close to the Fermi level. It was further reported that the Hall resistance shows a giant nonlinear behavior originating from a series of magnetic-field-induced Lifshitz transitions in the spin-dependent band structure (Fig. 12h) [226]. Combined with band structure calculation, these results suggest that in EuCd$_2$As$_2$, electronic structure is extremely sensitive to the spin canting angle, with the magnetic field causing band crossing and band inversion and introducing a band gap when oriented along specific directions, offering an ideal platform for Berry curvature engineering.

## 4. Other magnetic topological metals

As introduced in the previous section, intrinsic magnetic TIs have nontrivial bulk band topology featured by a global bulk gap and TSS residing inside the bulk gap. Chemical potential can be tuned into the bulk gap to eliminate the transport contribution from the bulk bands, a key prerequisite to realize quantized Hall transport. There exist other magnetic systems which lack a global bulk gap in the whole momentum space but possess a locally nontrivial bulk gap and TSS inside. Such systems always exhibit metallic transport behavior contributed by trivial bulk bands. Anomalous Hall effect (AHE) is generally expected from the coexisting net magnetic moment and locally nontrivial topology. We term such materials as magnetic topological metals. It is noted that there is no strict theoretical scheme describing magnetic topological metal since the metallicity doesn't only come from band-topology-induced TSS but rather the trivial bulk bands. We choose this term only to emphasize its distinction from intrinsic magnetic TIs and topological SMs.

### a) $Fe_3Sn_2$

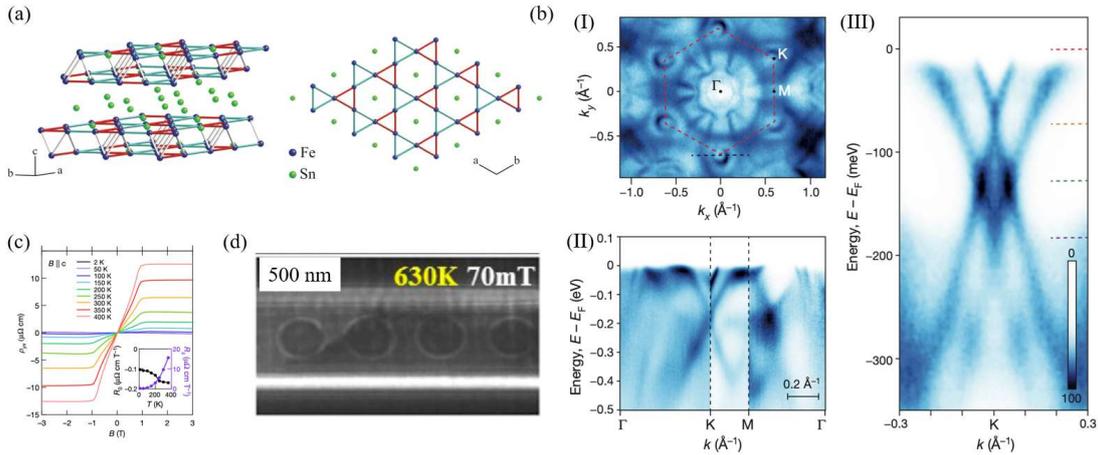

FIG. 13. (a) Crystal structure schematic of $Fe_3Sn_2$, adapted from [235]. (b) Fermi surface (I), high symmetry line band structure (II) and gapped Dirac cones at $K$ point (III), from [236]. (c) Field dependent Hall resistivity and the extracted ordinary and anomalous Hall coefficients, from [236]. (d) Under-focused Lorentz transmission electron microscopy images of skyrmionic bubbles in the 600 nm nanostripe taken at temperature $630\ K$ with magnetic field $70\ mT$.

$Fe_3Sn_2$ is a layered Kagome compound with a space group of $R\bar{3}m$ formed by interlacing two $Fe_3Sn$ layers and one Sn layer. The Fe atoms in the $Fe_3Sn$ layer form a Kagome structure, and the Sn atoms exhibit a honeycomb structure. The Sn atomic layer also exhibits a honeycomb distribution (Fig. 13a) [237]. $Fe_3Sn_2$ is ferromagnetic in the ground state with a Curie temperature of $T_C \sim 610K$ [235,238-240]. Due to the weak binding force between layers, $Fe_3Sn_2$ produces three different cleavage planes, $Fe_3Sn$-1-termination, $Fe_3Sn$-2-termination, and Sn-termination [236,241]. The experimentally observed band structures mainly come from $Fe_3Sn$-1-termination. The shape of the Fermi surface confirms the trigonal structure of $Fe_3Sn_2$. ARPES measurements have revealed two Dirac cone features at the corner of BZ, which are gapped by the SOC effect (Fig. 13b). Such strong SOC also couples the magnetic and electronic structure of Kagome lattice, exhibiting a magnetization-driven giant nematic (two-fold-symmetric) energy shift [242]. In the Kagome lattice, the destructive interference of the electron Bloch wave function can effectively localize the electrons to produce flat bands. Such flat bands are observed in $Fe_3Sn_2$, which are ~ 0.2 eV below the Fermi level [241].

The coexistence of nontrivial band topology and FM order in $Fe_3Sn_2$ produces giant AHE [40,236,243]. The measured anomalous Hall conductivity (AHC) is found to be temperature independent and persists above room temperature (Fig. 13c), which is suggestive of prominent Berry curvature from the time-reversal-symmetry-breaking electronic bands of the Kagome plane. Moreover, $Fe_3Sn_2$ shows complex magnetic bubbles and magnetic vortex structure like skyrmions [244-249]. These bubbles are three-dimensional magnetic domains with complicated evolution of spin texture, which not only give rise to topological Hall transport response, but also show record-high temperature stability in magnetic racetrack memory devices (Fig. 13d).

**b)   $RT_6X_6$ ($R$=Rare earth metal; T=transition metal; X=Sn, Ge)**

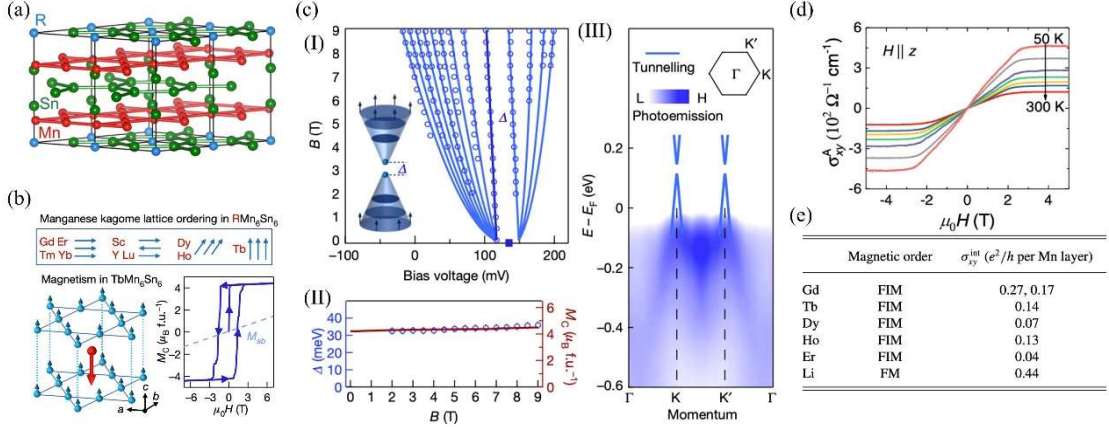

FIG.14 (a) $R$Mn$_6$Sn$_6$ lattice structure comprised of different layers of Mn$_3$Sn, $R$Sn, and Sn atoms, from [250]. (b) Magnetic structure of $R$Mn$_6$Sn$_6$ with the direction of magnetic moments depending on the $R$ site element, from [251] (c) Fitting the Landau fan data from field dependent STS measurements on TbMn$_6$Se$_6$ (open circles) with the spin polarized and Chern gapped Dirac dispersion (solid lines) (I) resulting field dependent size of the Dirac gap (II). Such gap is located above the Fermi level as indicated by the ARPES spectra in (III), from [251]. (d) Temperature dependent AHC of LiMn$_6$Sn$_6$ for magnetic field parallel to the $z$ axis, from [252]. (e) Comparison of the intrinsic AHC of LiMn$_6$Sn$_6$ with those of $R$Mn$_6$Sn$_6$, where FIM denotes the ferrimagnet and FM is the ferromagnet, from [252].

Layered Kagome compounds $RT_6X_6$ ($R$=rare earth metal, T=transition, alkali, alkaline earth metal, X=Sn or Ge) crystallize in the $P6/mmm$ space group. As shown in Fig. 14a, T$_3$X is the Kagome layer of T ions with one X atom at the center of the hexagon. In $R$X layer, the $R$ atom lies at the center of the hexagons surrounded by the X atoms. X layer is a hexagonal layer only consisting of X atoms and separating each unit cell. In this system, the 4f electrons in the $R$ element interact with the 3d electrons in the transition metal element T to generate a rich magnetic structure. Many novel physical properties are also found in this system, such as flat band, giant AHE and Nernst effects, etc. Recent published articles focus mostly on $R$Mn$_6$Sn$_6$ and $R$V$_6$Sn$_6$. Therefore, the following content will discuss these two systems.

Since Mn is a well-known magnetic metal, there are many magnetic configurations emerged due to the interaction between Mn 3d magnetic moment and $R$ 4f magnetic moments (Fig. 14b and 14e) [250,253-256]. When $R$ is a lanthanide element ($R$ = Gd-Tm, Lu), its magnetic configuration varies from ferrimagnetic to antiferromagnetic. For $R$ = Gd to Ho, their magnetic configuration is ferrimagnetic, and for $R$ = Er, Tm and Lu, they possess antiferromagnetic ground state. The direction of the magnetic moment of the $R$ element tends to be antiparallel to the magnetic moment of Mn, and the moment direction is variable for different $R$ elements. GdMn$_6$Sn$_6$ moment is arranged in-plane, and TbMn$_6$Sn$_6$ moment is arranged out-of-plane. DyMn$_6$Sn$_6$ and HoMn$_6$Sn$_6$ possess a conical magnetic structure. When $R$ is Er and Tm, the Mn and Er = Tm sublattices are independently ordered in an AFM manner because the strength of the magnetic coupling is weak. Since there is no 4f electrons in Lu and Y, they form in-plane FM and helical AFM along $c$-axis. For $R$ = Gd to Ho, the Curie temperature of them is 435, 423, 393, and 376 K, respectively. For $R$ = Er to Lu and Y, the Neel temperature of them is 352, 347, 353, and 333 K, respectively. In general, the electronic structure is closely related to magnetic configuration, when magnetic configuration change, the electronic structure will also change. However, for the $R$Mn$_6$Sn$_6$ ($R$ = Gd-Tm, Lu, Y) system, even for the different $R$, the band structure does not change significantly, indicating weak coupling between the low energy bands and magnetic moments.

Kagome lattice usually hosts three typical band features: flat band over the whole BZ, Dirac cones located at the BZ corners, and the saddle points located at the BZ boundary. Such features have indeed been observed in YMn$_6$Sn$_6$ and others by ARPES [257,258]. The strong correlation between magnetism and Kagome lattice can produce many novel physical properties. In TbMn$_6$Sn$_6$, its Kagome lattice features an out-of-plane magnetic ground state, so it is predicted to support the intrinsic Chern topological phase. In STM measurement, the Dirac cone with a Chern gap (Fig. 14c) and topological edge state are detected, implying its non-trivial topological nature [251].

The coexistence of nontrivial band topology and variation of magnetic structure results in novel transport behavior. In YMn$_6$Sn$_6$, a large room temperature anomalous transverse thermoelectric effect of $\approx 2~\mu V~K^{-1}$ is realized, larger than all canted AFM material studied to date at the room temperature [259]. In addition, topological Hall effect is observed in the transverse conical spiral phase of YMn$_6$Sn$_6$ and ErMn6Sn6 with similar magnetic configuration [260-262]. Large anomalous Hall conductivity is also observed in many $R$Mn$_6$Sn$_6$ compounds such as LiMn$_6$Sn$_6$, TbMn$_6$Sn$_6$, DyMn$_6$Sn$_6$, and HoMn$_6$Sn$_6$, as shown in Fig.14d-e [250,252,256,260,262].

In isostructural $R$V$_6$Sn$_6$ compounds, V atoms have no magnetic moments, so that $R$V$_6$Sn$_6$ magnetic configuration is different from $R$Mn$_6$Sn$_6$. The magnetic configuration is determined to be out-of-plane AFM for $R$ = Tb-Ho and in-plane AFM for $R$ = Er and Tm. because Lu and Y also possess no magnetic moment, so the compounds for $R$ = Lu and Y are paramagnetic metals [263]. Typical band features such as Dirac cone, saddle point, and flat bands are also observed in this family [264]. Furthermore, TSS Dirac cones emerge from the nontrivial bulk band topology and can be tuned in binding energy via potassium deposition [265].

c) EuAs$_3$

EuAs$_3$ crystallizes in a monoclinic structure (space group $C2/m$, No. 12). As shown in Fig. 15a, the moments of Eu are oriented along with $b$ axis [266]. The specific heat, electrical conductivity, susceptibility measurements [267], neutron diffraction [268], X-ray scattering technique [269,270] and μSR [271] studies showed that EuAs$_3$ orders in an incommensurate antiferromagnetic state at $T_N = 11\ K$, and goes through an incommensurate-commensurate lock-in phase transition at $T_L = 10.3\ K$, reaching a collinear antiferromagnetic ground state. Electrical transport studies showed an extremely anisotropic magnetoresistance related to the magnetic configuration [272].

Recently, the magnetism-induced topology of EuAs$_3$ has been demonstrated and the origin of extremely anisotropic magnetoresistance has been discussed [266]. An unsaturated extremely anisotropic magnetoresistance of ~2×10$^5$% at 1.8 $K$ and 28.3 $T$ has been observed, as shown in Fig. 15b. Meanwhile, through the DFT calculations and transport measurements, it is demonstrated that EuAs$_3$ is a magnetic topological massive Dirac metal at AFM ground state. ARPES results probed by different photon energies verify that EuAs$_3$ is a topological nodal semimetal in paramagnetic state (Fig. 15d-e), this is related to the extremely anisotropic magnetoresistance. For 3≤T≤30 K, the concentration of hole carriers is larger than that of electron carriers. Upon decreasing the temperature $T < 3\ K$ the concentration of electron carriers is suddenly enhanced, accompanied by a sharp increase in the mobility of hole carriers, indicating a possible Lifshitz transition (Fig. 15c).

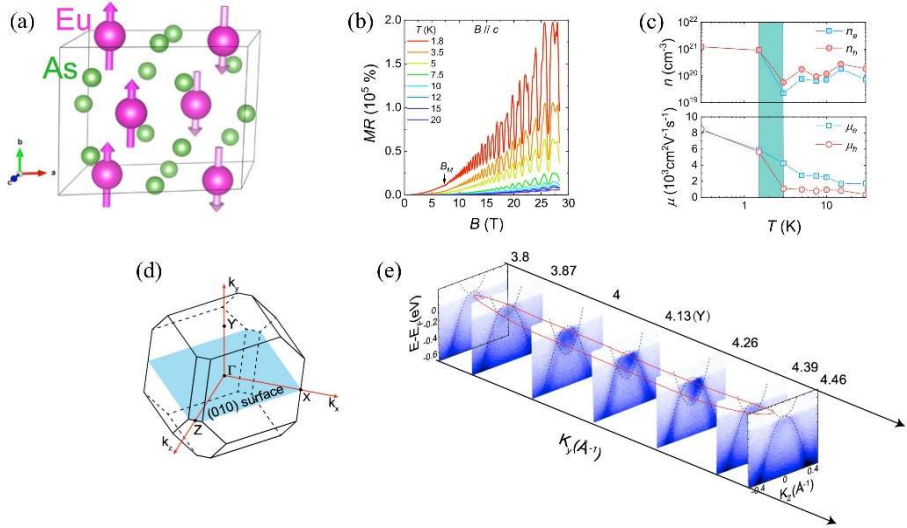

FIG. 15. (a) crystal structure of EuAs$_3$ [266]. (b) Magnetoresistance measurements [266]. (c) Carrier concentration and mobility [266]. (d) The Brillouin zone of EuAs$_3$, with high-symmetry points and (010) surface labeled [266]. (e) The band dispersions along $k_y$ direction probed by different photon energies. The red ellipse illustrates the topological nontrivial nodal loop schematically [266].

## 5. Perspective

In this review, we have gone through several intrinsic magnetic topological states of matter by introducing their representing materials. The interaction between magnetic order and band topology in these materials brings forth characteristic band features such as Dirac gap, Weyl point, Fermi arc, hinge/corner state and so on, produces large Berry curvature and enables novel topological transport responses including quantum anomalous Hall effect, intrinsic anomalous Hall effect, anomalous Nernst effect, negative magnetoresistance as the signature of chiral anomaly and so on. Intrinsic magnetic topological insulators are of fundamental and practical importance because of the potential for the development of dissipationless spintronics, information storage and quantum computation. However, so far only $Mn(Bi,Sb)_2Te_4 \cdot ((Bi,Sb)_2Te_3)_n$ family is firmly verified as intrinsic magnetic topological insulator. For this family of materials, the lack of sizable magnetic gap hinders the realization of quantum anomalous Hall effect at the expected temperature. It is thus highly desired to search for new material systems hosting such topological state. Instead of incorporating magnetism into established topological systems like the way how $Mn(Bi,Sb)_2Te_4 \cdot ((Bi,Sb)_2Te_3)_n$ and magnetically doped $Bi_2(Se,Te)_3$ families are realized, we envision that looking for band topology based on known ferromagnets or antiferromagnets will be more efficient to realize intrinsic magnetic topological insulator.

While we are concentrating on the interplay between magnetism and band topology in these quantum states of matter, it is well known that magnets host many ordered phases such as spin density wave, charge density wave, superconductivity, nematicity and so on. The interplay between band topology and these orders could generate exotic states such as axionic charge-density wave [273], chiral Majorana fermions [274] and the unknown which deserved future theoretical and experimental investigation.

**ACKNOWLEDGEMENTS**

This work is supported by the National Key R&D Program of China (Grants No. 2022YFA1403700 and 2020YFA0308900), the National Natural Science Foundation of China (NSFC) (Grants No. 12074163, 12074161, 11504159), NSFC Guangdong (No. 2016A030313650), Guangdong Basic and Applied Basic Research Foundation (Grants No. 2022B1515020046, 2022B1515130005 and 2021B1515130007), the Guangdong Innovative and Entrepreneurial Research Team Program (Grants No. 2019ZT08C044, 2016ZT06D348), Shenzhen Science and Technology Program (Grant No. KQTD20190929173815000). C.C. acknowledges the assistance of SUSTech Core Research Facilities. C. L. acknowledges additional support from the Highlight Project (No. PHYS-HL-2020-1) of the College of Science, SUSTech.

[1]     C. L. Kane & E. J. Mele. Z2 topological order and the quantum spin Hall effect. *Physical Review Letters* **95**, 146802, doi:10.1103/PhysRevLett.95.146802 (2005).

[2]     L. Fu & C. L. Kane. Topological insulators with inversion symmetry. *Physical Review B* **76**, 045302 (2007).

[3]     L. Fu, C. L. Kane & E. J. Mele. Topological insulators in three dimensions. *Physical Review Letters* **98**, - (2007).

[4]     J. E. Moore & L. Balents. Topological invariants of time-reversal-invariant band structures. *Physical Review B* **75**, doi:10.1103/PhysRevB.75.121306 (2007).

[5]     H. J. Zhang, C. X. Liu, X. L. Qi, X. Dai, Z. Fang & S. C. Zhang. Topological insulators in Bi2Se3, Bi2Te3 and Sb2Te3 with a single Dirac cone on the surface. *Nature Physics* **5**, 438-442 (2009).

[6]     B. A. Bernevig, T. L. Hughes & S. C. Zhang. Quantum spin Hall effect and topological phase transition in HgTe quantum wells. *Science* **314**, 1757-1761, doi:DOI 10.1126/science.1133734 (2006).

[7]     M. Konig, S. Wiedmann, C. Brune, A. Roth, H. Buhmann, L. W. Molenkamp, X. L. Qi & S. C. Zhang. Quantum spin hall insulator state in HgTe quantum wells. *Science* **318**, 766-770, doi:DOI 10.1126/science.1148047 (2007).

[8]     D. Hsieh, D. Qian, L. Wray, Y. Xia, Y. S. Hor, R. J. Cava & M. Z. Hasan. A topological Dirac insulator in a quantum spin Hall phase. *Nature* **452**, 970-974 (2008).

[9]     Y. L. Chen, J. G. Analytis, J. H. Chu, Z. K. Liu, S. K. Mo, X. L. Qi, H. J. Zhang, D. H. Lu, X. Dai, Z. Fang, S. C. Zhang, I. R. Fisher, Z. Hussain & Z. X. Shen. Experimental Realization of a Three-Dimensional Topological Insulator, Bi2Te3. *Science* **325**, 178-181 (2009).

[10]    D. Hsieh, Y. Xia, D. Qian, L. Wray, J. H. Dil, F. Meier, J. Osterwalder, L. Patthey, J. G. Checkelsky, N. P. Ong, A. V. Fedorov, H. Lin, A. Bansil, D. Grauer, Y. S. Hor, R. J. Cava & M. Z. Hasan. A tunable topological insulator in the spin helical Dirac transport regime. *Nature* **460**, 1101-U1159 (2009).

[11]    Y. Xia, D. Qian, D. Hsieh, L. Wray, A. Pal, H. Lin, A. Bansil, D. Grauer, Y. S. Hor, R. J. Cava & M. Z. Hasan. Observation of a large-gap topological-insulator class with a single Dirac cone on the surface. *Nature Physics* **5**, 398-402 (2009).

[12]    T. Valla, Z. H. Pan, D. Gardner, Y. S. Lee & S. Chu. Photoemission Spectroscopy of Magnetic and Nonmagnetic Impurities on the Surface of the Bi_{2}Se_{3} Topological Insulator. *Physical Review Letters* **108**, 117601 (2012).

[13]    Chaoyu Chen, Shaolong He, Hongming Weng, Wentao Zhang, Lin Zhao, Haiyun Liu, Xiaowen Jia, Daixiang Mou, Shanyu Liu, Junfeng He, Yingying Peng, Ya Feng, Zhuojin Xie, Guodong Liu, Xiaoli Dong, Jun Zhang, Xiaoyang Wang, Qinjun Peng, Zhimin Wang, Shenjin Zhang, Feng Yang, Chuangtian Chen, Zuyan Xu, Xi Dai, Zhong Fang & X. J. Zhou. Robustness of topological order and formation of quantum well states in topological insulators exposed to ambient environment. *Proceedings of the National Academy of Sciences* **109**, 3694-3698, doi:10.1073/pnas.1115555109 (2012).

[14]    L. Andrew Wray, Su-Yang Xu, Yuqi Xia, David Hsieh, Alexei V. Fedorov, Yew San Hor, Robert J. Cava, Arun Bansil, Hsin Lin & M. Zahid Hasan. A topological insulator surface under strong Coulomb, magnetic and disorder perturbations. *Nature Physics* **7**, 32-37, doi:http://www.nature.com/nphys/journal/v7/n1/abs/nphys1838.html#supplementary-information (2011).

[15] Cui-Zu Chang, Jinsong Zhang, Xiao Feng, Jie Shen, Zuocheng Zhang, Minghua Guo, Kang Li, Yunbo Ou, Pang Wei, Li-Li Wang, Zhong-Qing Ji, Yang Feng, Shuaihua Ji, Xi Chen, Jinfeng Jia, Xi Dai, Zhong Fang, Shou-Cheng Zhang, Ke He, Yayu Wang, Li Lu, Xu-Cun Ma & Qi-Kun Xue. Experimental Observation of the Quantum Anomalous Hall Effect in a Magnetic Topological Insulator. *Science*, doi:10.1126/science.1234414 (2013).

[16] Rui Yu, Wei Zhang, Hai-Jun Zhang, Shou-Cheng Zhang, Xi Dai & Zhong Fang. Quantized Anomalous Hall Effect in Magnetic Topological Insulators. *Science* **329**, 61-64, doi:10.1126/science.1187485 (2010).

[17] N. P Armitage, E. J Mele & Ashvin Vishwanath. Weyl and Dirac semimetals in three-dimensional solids. *Reviews of Modern Physics* **90**, doi:10.1103/RevModPhys.90.015001 (2018).

[18] A. A. Burkov & Leon Balents. Weyl Semimetal in a Topological Insulator Multilayer. *Physical Review Letters* **107**, 127205 (2011).

[19] A. A. Burkov, M. D. Hook & Leon Balents. Topological nodal semimetals. *Physical Review B* **84**, 235126 (2011).

[20] X. G. Wan, A. M. Turner, A. Vishwanath & S. Y. Savrasov. Topological semimetal and Fermi-arc surface states in the electronic structure of pyrochlore iridates. *Physical Review B* **83** (2011).

[21] Zhijun Wang, Yan Sun, Xing-Qiu Chen, Cesare Franchini, Gang Xu, Hongming Weng, Xi Dai & Zhong Fang. Dirac semimetal and topological phase transitions in A3Bi (A=Na, K, Rb). *Physical Review B* **85**, 195320 (2012).

[22] S. M. Young, S. Zaheer, J. C. Y. Teo, C. L. Kane, E. J. Mele & A. M. Rappe. Dirac Semimetal in Three Dimensions. *Physical Review Letters* **108**, 140405 (2012).

[23] Z. J. Wang, H. M. Weng, Q. S. Wu, X. Dai & Z. Fang. Three-dimensional Dirac semimetal and quantum transport in Cd3As2. *Physical Review B* **88**, doi:10.1103/PhysRevB.88.125427 (2013).

[24] Wei Ning & Zhiqiang Mao. Recent advancements in the study of intrinsic magnetic topological insulators and magnetic Weyl semimetals. *APL Materials* **8**, doi:10.1063/5.0015328 (2020).

[25] M. Zahid Hasan, Guoqing Chang, Ilya Belopolski, Guang Bian, Su-Yang Xu & Jia-Xin Yin. Weyl, Dirac and high-fold chiral fermions in topological quantum matter. *Nature Reviews Materials* **6**, 784-803, doi:10.1038/s41578-021-00301-3 (2021).

[26] B. A. Bernevig, C. Felser & H. Beidenkopf. Progress and prospects in magnetic topological materials. *Nature* **603**, 41-51, doi:10.1038/s41586-021-04105-x (2022).

[27] F. Tang, H. C. Po, A. Vishwanath & X. Wan. Comprehensive search for topological materials using symmetry indicators. *Nature* **566**, 486-489, doi:10.1038/s41586-019-0937-5 (2019).

[28] M. G. Vergniory, L. Elcoro, C. Felser, N. Regnault, B. A. Bernevig & Z. Wang. A complete catalogue of high-quality topological materials. *Nature* **566**, 480-485, doi:10.1038/s41586-019-0954-4 (2019).

[29] T. Zhang, Y. Jiang, Z. Song, H. Huang, Y. He, Z. Fang, H. Weng & C. Fang. Catalogue of topological electronic materials. *Nature* **566**, 475-479, doi:10.1038/s41586-019-0944-6 (2019).

[30] 刘畅 & 刘祥瑞. 强三维拓扑绝缘体与磁性拓扑绝缘体的角分辨光电子能谱学研究进展. *物理学报* **68**, 227901, doi:- 10.7498/aps.68.20191450 (2019).

[31] Yoichi Ando. Topological Insulator Materials. *Journal of the Physical Society of Japan* **82**, doi:10.7566/jpsj.82.102001 (2013).

[32] Jonathan A. Sobota, Yu He & Zhi-Xun Shen. Angle-resolved photoemission studies of quantum materials. *Reviews of Modern Physics* **93**, 025006, doi:10.1103/RevModPhys.93.025006 (2021).

[33] Yuan Wang. On the topological surface states of the intrinsic magnetic topological insulator Mn-Bi-Te family. *arXiv:2211.04017* (2022).

[34] Yufei Zhao & Qihang Liu. Routes to realize the axion-insulator phase in MnBi2Te4(Bi2Te3)n family. *Applied Physics Letters* **119**, doi:10.1063/5.0059447 (2021).

[35] P. Wang, J. Ge, J. Li, Y. Liu, Y. Xu & J. Wang. Intrinsic magnetic topological insulators. *Innovation* **2**, 100098, doi:10.1016/j.xinn.2021.100098 (2021).

[36] Yang Li & Yong Xu. First-principles discovery of novel quantum physics and materials: From theory to experiment. *Computational Materials Science* **190**, 110262, doi:10.1016/j.commatsci.2020.110262 (2021).

[37] 陈朝宇. 磁性起源的表面态能隙与"半磁拓扑绝缘体". *物理* **50**, 267-267 (2021).

[38] Ke He. MnBi2Te4-family intrinsic magnetic topological materials. *npj Quantum Materials* **5**, 90, doi:10.1038/s41535-020-00291-5 (2020).

[39] 占国慧,王怀强,张海军. 反铁磁拓扑绝缘体与轴子绝缘体-MnBi2Te4 系列磁性体系的研究进展. *物理* **49**, 817-827, doi:10.7693/wl20201203 (2020).

[40] T. Kida, L. A. Fenner, A. A. Dee, I. Terasaki, M. Hagiwara & A. S. Wills. The giant anomalous Hall effect in the ferromagnet Fe3Sn2--a frustrated kagome metal. *J Phys Condens Matter* **23**, 112205, doi:10.1088/0953-8984/23/11/112205 (2011).

[41] F. D. M. Haldane. Model for a Quantum Hall Effect without Landau Levels: Condensed-Matter Realization of the "Parity Anomaly". *Physical Review Letters* **61**, 2015-2018 (1988).

[42] Po-Yao Chang, Onur Erten & Piers Coleman. Möbius Kondo insulators. *Nature Physics* **13**, 794-798, doi:10.1038/nphys4092 (2017).

[43] Ken Shiozaki, Masatoshi Sato & Kiyonori Gomi. Z2topology in nonsymmorphic crystalline insulators: Möbius twist in surface states. *Physical Review B* **91**, doi:10.1103/PhysRevB.91.155120 (2015).

[44] R. X. Zhang, F. Wu & S. Das Sarma. Mobius Insulator and Higher-Order Topology in MnBi2nTe3n+1. *Phys Rev Lett* **124**, 136407, doi:10.1103/PhysRevLett.124.136407 (2020).

[45] Chang Liu, Yongchao Wang, Hao Li, Yang Wu, Yaoxin Li, Jiaheng Li, Ke He, Yong Xu, Jinsong Zhang & Yayu Wang. Robust axion insulator and Chern insulator phases in a two-dimensional antiferromagnetic topological insulator. *Nature Materials* **19**, 522-527, doi:10.1038/s41563-019-0573-3 (2020).

[46] Cui-Zu Chang, Jinsong Zhang, Xiao Feng, Jie Shen, Zuocheng Zhang, Minghua Guo, Kang Li, Yunbo Ou, Pang Wei, Li-Li Wang, Zhong-Qing Ji, Yang Feng, Shuaihua Ji, Xi Chen, Jinfeng Jia, Xi Dai, Zhong Fang, Shou-Cheng Zhang, Ke He, Yayu Wang, Li Lu, Xu-Cun Ma & Qi-Kun Xue. Experimental Observation of the Quantum Anomalous Hall Effect in a Magnetic Topological Insulator. *Science* **340**, 167, doi:10.1126/science.1234414 (2013).

[47] Liang Wu, M. Salehi, N. Koirala, J. Moon, S. Oh & N. P. Armitage. Quantized Faraday and Kerr rotation and axion electrodynamics of a 3D topological insulator. *Science* **354**, 1124-1127, doi:10.1126/science.aaf5541 (2016).

[48] E. Liu, Y. Sun, N. Kumar, L. Muchler, A. Sun, L. Jiao, S. Y. Yang, D. Liu, A. Liang, Q. Xu, J.

Kroder, V. Suss, H. Borrmann, C. Shekhar, Z. Wang, C. Xi, W. Wang, W. Schnelle, S. Wirth, Y. Chen, S. T. B. Goennenwein & C. Felser. Giant anomalous Hall effect in a ferromagnetic Kagome-lattice semimetal. *Nat Phys* **14**, 1125-1131, doi:10.1038/s41567-018-0234-5 (2018).

[49] C. L. Zhang, S. Y. Xu, I. Belopolski, Z. Yuan, Z. Lin, B. Tong, G. Bian, N. Alidoust, C. C. Lee, S. M. Huang, T. R. Chang, G. Chang, C. H. Hsu, H. T. Jeng, M. Neupane, D. S. Sanchez, H. Zheng, J. Wang, H. Lin, C. Zhang, H. Z. Lu, S. Q. Shen, T. Neupert, M. Zahid Hasan & S. Jia. Signatures of the Adler-Bell-Jackiw chiral anomaly in a Weyl fermion semimetal. *Nat Commun* **7**, 10735, doi:10.1038/ncomms10735 (2016).

[50] S. N. Guin, P. Vir, Y. Zhang, N. Kumar, S. J. Watzman, C. Fu, E. Liu, K. Manna, W. Schnelle, J. Gooth, C. Shekhar, Y. Sun & C. Felser. Zero-Field Nernst Effect in a Ferromagnetic Kagome-Lattice Weyl-Semimetal $Co_3Sn_2S_2$. *Adv Mater* **31**, e1806622, doi:10.1002/adma.201806622 (2019).

[51] EnKe Liu & Shen Zhang. Topologically enhanced zero-field transverse Nernst thermoelectric effect in magnetic topological semimetals. *SCIENTIA SINICA Physica, Mechanica & Astronomica* **49**, doi:10.1360/sspma-2019-0367 (2019).

[52] Roger S. K. Mong, Andrew M. Essin & Joel E. Moore. Antiferromagnetic topological insulators. *Physical Review B* **81**, doi:10.1103/PhysRevB.81.245209 (2010).

[53] M. M. Otrokov, T. V. Menshchikova, M. G. Vergniory, I. P. Rusinov, A. Yu Vyazovskaya, Yu M. Koroteev, G. Bihlmayer, A. Ernst, P. M. Echenique, A. Arnau & E. V. Chulkov. Highly-ordered wide bandgap materials for quantized anomalous Hall and magnetoelectric effects. *2D Materials* **4**, doi:10.1088/2053-1583/aa6bec (2017).

[54] M. M. Otrokov, T. V. Menshchikova, I. P. Rusinov, M. G. Vergniory, V. M. Kuznetsov & E. V. Chulkov. Magnetic extension as an efficient method for realizing the quantum anomalous hall state in topological insulators. *JETP Letters* **105**, 297-302, doi:10.1134/s0021364017050113 (2017).

[55] Yan Gong, Jingwen Guo, Jiaheng Li, Kejing Zhu, Menghan Liao, Xiaozhi Liu, Qinghua Zhang, Lin Gu, Lin Tang, Xiao Feng, Ding Zhang, Wei Li, Canli Song, Lili Wang, Pu Yu, Xi Chen, Yayu Wang, Hong Yao, Wenhui Duan, Yong Xu, Shou-Cheng Zhang, Xucun Ma, Qi-Kun Xue & Ke He. Experimental Realization of an Intrinsic Magnetic Topological Insulator. *Chinese Physics Letters* **36**, doi:10.1088/0256-307x/36/7/076801 (2019).

[56] D. Zhang, M. Shi, T. Zhu, D. Xing, H. Zhang & J. Wang. Topological Axion States in the Magnetic Insulator $MnBi_2Te_4$ with the Quantized Magnetoelectric Effect. *Phys Rev Lett* **122**, 206401, doi:10.1103/PhysRevLett.122.206401 (2019).

[57] M. M. Otrokov, I. P. Rusinov, M. Blanco-Rey, M. Hoffmann, A. Y. Vyazovskaya, S. V. Eremeev, A. Ernst, P. M. Echenique, A. Arnau & E. V. Chulkov. Unique Thickness-Dependent Properties of the van der Waals Interlayer Antiferromagnet $MnBi_2Te_4$ Films. *Phys Rev Lett* **122**, 107202, doi:10.1103/PhysRevLett.122.107202 (2019).

[58] Jiaheng Li, Yang Li, Shiqiao Du, Zun Wang, Bing-Lin Gu, Shou-Cheng Zhang, Ke He, Wenhui Duan & Yong Xu. Intrinsic magnetic topological insulators in van der Waals layered $MnBi_2Te_4$-family materials. *Science Advances* **5**, eaaw5685, doi:10.1126/sciadv.aaw5685 (2019).

[59] M. M. Otrokov, Klimovskikh, II, H. Bentmann, D. Estyunin, A. Zeugner, Z. S. Aliev, S. Gass, A. U. B. Wolter, A. V. Koroleva, A. M. Shikin, M. Blanco-Rey, M. Hoffmann, I. P. Rusinov, A.

Y. Vyazovskaya, S. V. Eremeev, Y. M. Koroteev, V. M. Kuznetsov, F. Freyse, J. Sanchez-Barriga, I. R. Amiraslanov, M. B. Babanly, N. T. Mamedov, N. A. Abdullayev, V. N. Zverev, A. Alfonsov, V. Kataev, B. Buchner, E. F. Schwier, S. Kumar, A. Kimura, L. Petaccia, G. Di Santo, R. C. Vidal, S. Schatz, K. Kissner, M. Unzelmann, C. H. Min, S. Moser, T. R. F. Peixoto, F. Reinert, A. Ernst, P. M. Echenique, A. Isaeva & E. V. Chulkov. Prediction and observation of an antiferromagnetic topological insulator. *Nature* **576**, 416-422, doi:10.1038/s41586-019-1840-9 (2019).

[60] Ziya S. Aliev, Imamaddin R. Amiraslanov, Daria I. Nasonova, Andrei V. Shevelkov, Nadir A. Abdullayev, Zakir A. Jahangirli, Elnur N. Orujlu, Mikhail M. Otrokov, Nazim T. Mamedov, Mahammad B. Babanly & Evgueni V. Chulkov. Novel ternary layered manganese bismuth tellurides of the MnTe-Bi2Te3 system: Synthesis and crystal structure. *Journal of Alloys and Compounds* **789**, 443-450, doi:10.1016/j.jallcom.2019.03.030 (2019).

[61] J. Wu, F. Liu, C. Liu, Y. Wang, C. Li, Y. Lu, S. Matsuishi & H. Hosono. Toward 2D Magnets in the (MnBi2Te4)(Bi2Te3)n Bulk Crystal. *Adv Mater* **32**, e2001815, doi:10.1002/adma.202001815 (2020).

[62] Daniel Souchay, Markus Nentwig, Daniel Günther, Simon Keilholz, Johannes de Boor, Alexander Zeugner, Anna Isaeva, Michael Ruck, Anja U. B. Wolter, Bernd Büchner & Oliver Oeckler. Layered manganese bismuth tellurides with GeBi4Te7- and GeBi6Te10-type structures: towards multifunctional materials. *Journal of Materials Chemistry C* **7**, 9939-9953, doi:10.1039/c9tc00979e (2019).

[63] Dong Sun Lee, Tae-Hoon Kim, Cheol-Hee Park, Chan-Yeup Chung, Young Soo Lim, Won-Seon Seo & Hyung-Ho Park. Crystal structure, properties and nanostructuring of a new layered chalcogenide semiconductor, Bi2MnTe4. *CrystEngComm* **15**, 5532-5538 doi:10.1039/c3ce40643a (2013).

[64] T. Hirahara, S. V. Eremeev, T. Shirasawa, Y. Okuyama, T. Kubo, R. Nakanishi, R. Akiyama, A. Takayama, T. Hajiri, S. I. Ideta, M. Matsunami, K. Sumida, K. Miyamoto, Y. Takagi, K. Tanaka, T. Okuda, T. Yokoyama, S. I. Kimura, S. Hasegawa & E. V. Chulkov. Large-Gap Magnetic Topological Heterostructure Formed by Subsurface Incorporation of a Ferromagnetic Layer. *Nano Lett* **17**, 3493-3500, doi:10.1021/acs.nanolett.7b00560 (2017).

[65] Joseph A. Hagmann, Xiang Li, Sugata Chowdhury, Si-Ning Dong, Sergei Rouvimov, Sujitra J. Pookpanratana, Kin Man Yu, Tatyana A. Orlova, Trudy B. Bolin, Carlo U. Segre, David G. Seiler, Curt A. Richter, Xinyu Liu, Margaret Dobrowolska & Jacek K. Furdyna. Molecular beam epitaxy growth and structure of self-assembled Bi2Se3/Bi2MnSe4 multilayer heterostructures. *New Journal of Physics* **19**, doi:10.1088/1367-2630/aa759c (2017).

[66] Lei Ding, Chaowei Hu, Feng Ye, Erxi Feng, Ni Ni & Huibo Cao. Crystal and magnetic structures of magnetic topological insulators MnBi2Te4 and MnBi4Te7. *Physical Review B* **101**, doi:10.1103/PhysRevB.101.020412 (2020).

[67] J. Q. Yan, Q. Zhang, T. Heitmann, Zengle Huang, K. Y. Chen, J. G. Cheng, Weida Wu, D. Vaknin, B. C. Sales & R. J. McQueeney. Crystal growth and magnetic structure of MnBi2Te4. *Physical Review Materials* **3**, 064202, doi:10.1103/PhysRevMaterials.3.064202 (2019).

[68] M. Z. Shi, B. Lei, C. S. Zhu, D. H. Ma, J. H. Cui, Z. L. Sun, J. J. Ying & X. H. Chen. Magnetic and transport properties in the magnetic topological insulators MnBi2Te4(Bi2Te3)n(n=1,2). *Physical Review B* **100**, 155144, doi:10.1103/PhysRevB.100.155144 (2019).

[69] Yu-Jie Hao, Pengfei Liu, Yue Feng, Xiao-Ming Ma, Eike F. Schwier, Masashi Arita, Shiv Kumar, Chaowei Hu, Rui'e Lu, Meng Zeng, Yuan Wang, Zhanyang Hao, Hong-Yi Sun, Ke Zhang, Jiawei Mei, Ni Ni, Liusuo Wu, Kenya Shimada, Chaoyu Chen, Qihang Liu & Chang Liu. Gapless Surface Dirac Cone in Antiferromagnetic Topological Insulator MnBi2Te4. *Physical Review X* **9**, 041038, doi:10.1103/PhysRevX.9.041038 (2019).

[70] C. Hu, K. N. Gordon, P. Liu, J. Liu, X. Zhou, P. Hao, D. Narayan, E. Emmanouilidou, H. Sun, Y. Liu, H. Brawer, A. P. Ramirez, L. Ding, H. Cao, Q. Liu, D. Dessau & N. Ni. A van der Waals antiferromagnetic topological insulator with weak interlayer magnetic coupling. *Nat Commun* **11**, 97, doi:10.1038/s41467-019-13814-x (2020).

[71] Xiao-Ming Ma, Zhongjia Chen, Eike F. Schwier, Yang Zhang, Yu-Jie Hao, Shiv Kumar, Ruie Lu, Jifeng Shao, Yuanjun Jin, Meng Zeng, Xiang-Rui Liu, Zhanyang Hao, Ke Zhang, Wumiti Mansuer, Chunyao Song, Yuan Wang, Boyan Zhao, Cai Liu, Ke Deng, Jiawei Mei, Kenya Shimada, Yue Zhao, Xingjiang Zhou, Bing Shen, Wen Huang, Chang Liu, Hu Xu & Chaoyu Chen. Hybridization-induced gapped and gapless states on the surface of magnetic topological insulators. *Physical Review B* **102**, 245136, doi:10.1103/PhysRevB.102.245136 (2020).

[72] Ruie Lu, Hongyi Sun, Shiv Kumar, Yuan Wang, Mingqiang Gu, Meng Zeng, Yu-Jie Hao, Jiayu Li, Jifeng Shao, Xiao-Ming Ma, Zhanyang Hao, Ke Zhang, Wumiti Mansuer, Jiawei Mei, Yue Zhao, Cai Liu, Ke Deng, Wen Huang, Bing Shen, Kenya Shimada, Eike F Schwier, Chang Liu, Qihang Liu & Chaoyu Chen. Half-Magnetic Topological Insulator with Magnetization-Induced Dirac Gap at a Selected Surface. *Physical Review X* **11**, 011039, doi:10.1103/PhysRevX.11.011039 (2021).

[73] Y. J Chen, L. X Xu, J. H Li, Y. W Li, H. Y Wang, C. F Zhang, H. Li, Y. Wu, A. J Liang, C. Chen, S. W Jung, C. Cacho, Y. H Mao, S. Liu, M. X Wang, Y. F Guo, Y. Xu, Z. K Liu, L. X Yang & Y. L Chen. Topological Electronic Structure and Its Temperature Evolution in Antiferromagnetic Topological Insulator MnBi2Te4. *Physical Review X* **9**, 041040, doi:10.1103/PhysRevX.9.041040 (2019).

[74] Hang Li, Shun-Ye Gao, Shao-Feng Duan, Yuan-Feng Xu, Ke-Jia Zhu, Shang-Jie Tian, Jia-Cheng Gao, Wen-Hui Fan, Zhi-Cheng Rao, Jie-Rui Huang, Jia-Jun Li, Da-Yu Yan, Zheng-Tai Liu, Wan-Ling Liu, Yao-Bo Huang, Yu-Liang Li, Yi Liu, Guo-Bin Zhang, Peng Zhang, Takeshi Kondo, Shik Shin, He-Chang Lei, You-Guo Shi, Wen-Tao Zhang, Hong-Ming Weng, Tian Qian & Hong Ding. Dirac Surface States in Intrinsic Magnetic Topological Insulators EuSn2As2 and MnBi2nTe3n+1. *Physical Review X* **9**, 041039, doi:10.1103/PhysRevX.9.041039 (2019).

[75] A. Liang, C. Chen, H. Zheng, W. Xia, K. Huang, L. Wei, H. Yang, Y. Chen, X. Zhang, X. Xu, M. Wang, Y. Guo, L. Yang, Z. Liu & Y. Chen. Approaching a Minimal Topological Electronic Structure in Antiferromagnetic Topological Insulator MnBi2Te4 via Surface Modification. *Nano Lett* **22**, 4307-4314, doi:10.1021/acs.nanolett.1c04930 (2022).

[76] Runzhe Xu, Yunhe Bai, Jingsong Zhou, Jiaheng Li, Xu Gu, Na Qin, Zhongxu Yin, Xian Du, Qinqin Zhang, Wenxuan Zhao, Yidian Li, Yang Wu, Cui Ding, Lili Wang, Aiji Liang, Zhongkai Liu, Yong Xu, Xiao Feng, Ke He, Yulin Chen & Lexian Yang. Evolution of the Electronic Structure of Ultrathin MnBi2Te4 Films. *Nano Letters* **22**, 6320-6327, doi:10.1021/acs.nanolett.2c02034 (2022).

[77] R. C. Vidal, H. Bentmann, T. R. F. Peixoto, A. Zeugner, S. Moser, C. H. Min, S. Schatz, K.

Kißner, M. Ünzelmann, C. I. Fornari, H. B. Vasili, M. Valvidares, K. Sakamoto, D. Mondal, J. Fujii, I. Vobornik, S. Jung, C. Cacho, T. K. Kim, R. J. Koch, C. Jozwiak, A. Bostwick, J. D. Denlinger, E. Rotenberg, J. Buck, M. Hoesch, F. Diekmann, S. Rohlf, M. Kalläne, K. Rossnagel, M. M. Otrokov, E. V. Chulkov, M. Ruck, A. Isaeva & F. Reinert. Surface states and Rashba-type spin polarization in antiferromagnetic MnBi2Te4 (0001). *Physical Review B* **100**, doi:10.1103/PhysRevB.100.121104 (2019).

[78]  Seng Huat Lee, Yanglin Zhu, Yu Wang, Leixin Miao, Timothy Pillsbury, Hemian Yi, Susan Kempinger, Jin Hu, Colin A. Heikes, P. Quarterman, William Ratcliff, Julie A. Borchers, Heda Zhang, Xianglin Ke, David Graf, Nasim Alem, Cui-Zu Chang, Nitin Samarth & Zhiqiang Mao. Spin scattering and noncollinear spin structure-induced intrinsic anomalous Hall effect in antiferromagnetic topological insulator MnBi2Te4. *Physical Review Research* **1**, doi:10.1103/PhysRevResearch.1.012011 (2019).

[79]  Yong Hu, Lixuan Xu, Mengzhu Shi, Aiyun Luo, Shuting Peng, Z. Y. Wang, J. J. Ying, T. Wu, Z. K. Liu, C. F. Zhang, Y. L. Chen, G. Xu, X. H. Chen & J. F. He. Universal gapless Dirac cone and tunable topological states in (MnBi2Te4)m(Bi2Te3)n heterostructures. *Physical Review B* **101**, 161113(R) (2020).

[80]  D. Nevola, H. X. Li, J. Q. Yan, R. G. Moore, H. N. Lee, H. Miao & P. D. Johnson. Coexistence of Surface Ferromagnetism and a Gapless Topological State in MnBi2Te4. *Phys Rev Lett* **125**, 117205, doi:10.1103/PhysRevLett.125.117205 (2020).

[81]  A. M. Shikin, D. A. Estyunin, Klimovskikh, II, S. O. Filnov, E. F. Schwier, S. Kumar, K. Miyamoto, T. Okuda, A. Kimura, K. Kuroda, K. Yaji, S. Shin, Y. Takeda, Y. Saitoh, Z. S. Aliev, N. T. Mamedov, I. R. Amiraslanov, M. B. Babanly, M. M. Otrokov, S. V. Eremeev & E. V. Chulkov. Nature of the Dirac gap modulation and surface magnetic interaction in axion antiferromagnetic topological insulator MnBi2Te4. *Sci Rep* **10**, 13226, doi:10.1038/s41598-020-70089-9 (2020).

[82]  Przemyslaw Swatek, Yun Wu, Lin-Lin Wang, Kyungchan Lee, Benjamin Schrunk, Jiaqiang Yan & Adam Kaminski. Gapless Dirac surface states in the antiferromagnetic topological insulator MnBi2Te4. *Physical Review B* **101**, 161109, doi:10.1103/PhysRevB.101.161109 (2020).

[83]  A. M. Shikin, D. A. Estyunin, N. L. Zaitsev, D. Glazkova, I. I. Klimovskikh, S. O. Filnov, A. G. Rybkin, E. F. Schwier, S. Kumar, A. Kimura, N. Mamedov, Z. Aliev, M. B. Babanly, K. Kokh, O. E. Tereshchenko, M. M. Otrokov, E. V. Chulkov, K. A. Zvezdin & A. K. Zvezdin. Sample-dependent Dirac-point gap in MnBi2Te4 and its response to applied surface charge: A combined photoemission and ab initio study. *Physical Review B* **104**, 115168, doi:10.1103/PhysRevB.104.115168 (2021).

[84]  R. C. Vidal, H. Bentmann, J. I. Facio, T. Heider, P. Kagerer, C. I. Fornari, T. R. F. Peixoto, T. Figgemeier, S. Jung, C. Cacho, B. Buchner, J. van den Brink, C. M. Schneider, L. Plucinski, E. F. Schwier, K. Shimada, M. Richter, A. Isaeva & F. Reinert. Orbital Complexity in Intrinsic Magnetic Topological Insulators MnBi4Te7 and MnBi6Te10. *Phys Rev Lett* **126**, 176403, doi:10.1103/PhysRevLett.126.176403 (2021).

[85]  Xuefeng Wu, Jiayu Li, Xiao-Ming Ma, Yu Zhang, Yuntian Liu, Chun-Sheng Zhou, Jifeng Shao, Qiaoming Wang, Yu-Jie Hao, Yue Feng, Eike F. Schwier, Shiv Kumar, Hongyi Sun, Pengfei Liu, Kenya Shimada, Koji Miyamoto, Taichi Okuda, Kedong Wang, Maohai Xie, Chaoyu Chen, Qihang Liu, Chang Liu & Yue Zhao. Distinct Topological Surface States on

the Two Terminations of MnBi4Te7. *Physical Review X* **10**, 031013, doi:10.1103/PhysRevX.10.031013 (2020).

[86] Shangjie Tian, Shunye Gao, Simin Nie, Yuting Qian, Chunsheng Gong, Yang Fu, Hang Li, Wenhui Fan, Peng Zhang, Takesh Kondo, Shik Shin, Johan Adell, Hanna Fedderwitz, Hong Ding, Zhijun Wang, Tian Qian & Hechang Lei. Magnetic topological insulator MnBi6Te10 with a zero-field ferromagnetic state and gapped Dirac surface states. *Physical Review B* **102**, doi:10.1103/PhysRevB.102.035144 (2020).

[87] Ilya I. Klimovskikh, Mikhail M. Otrokov, Dmitry Estyunin, Sergey V. Eremeev, Sergey O. Filnov, Alexandra Koroleva, Eugene Shevchenko, Vladimir Voroshnin, Artem G. Rybkin, Igor P. Rusinov, Maria Blanco-Rey, Martin Hoffmann, Ziya S. Aliev, Mahammad B. Babanly, Imamaddin R. Amiraslanov, Nadir A. Abdullayev, Vladimir N. Zverev, Akio Kimura, Oleg E. Tereshchenko, Konstantin A. Kokh, Luca Petaccia, Giovanni Di Santo, Arthur Ernst, Pedro M. Echenique, Nazim T. Mamedov, Alexander M. Shikin & Eugene V. Chulkov. Tunable 3D/2D magnetism in the (MnBi2Te4)(Bi2Te3)m topological insulators family. *npj Quantum Materials* **5**, doi:10.1038/s41535-020-00255-9 (2020).

[88] Na Hyun Jo, Lin-Lin Wang, Robert-Jan Slager, Jiaqiang Yan, Yun Wu, Kyungchan Lee, Benjamin Schrunk, Ashvin Vishwanath & Adam Kaminski. Intrinsic axion insulating behavior in antiferromagnetic MnBi6Te10. *Physical Review B* **102**, doi:10.1103/PhysRevB.102.045130 (2020).

[89] Chaowei Hu, Lei Ding, Kyle N. Gordon, Barun Ghosh, Hung-Ju Tien, Haoxiang Li, A. Garrison Linn, Shang-Wei Lien, Cheng-Yi Huang, Scott Mackey, Jinyu Liu, P. V. Sreenivasa Reddy, Bahadur Singh, Amit Agarwal, Arun Bansil, Miao Song, Dongsheng Li, Su-Yang Xu, Hsin Lin, Huibo Cao, Tay-Rong Chang, Dan Dessau & Ni Ni. Realization of an intrinsic ferromagnetic topological state in MnBi8Te13. *Science Advances* **6**, eaba4275, doi:10.1126/sciadv.aba4275 (2020).

[90] T. Hirahara, M. M. Otrokov, T. T. Sasaki, K. Sumida, Y. Tomohiro, S. Kusaka, Y. Okuyama, S. Ichinokura, M. Kobayashi, Y. Takeda, K. Amemiya, T. Shirasawa, S. Ideta, K. Miyamoto, K. Tanaka, S. Kuroda, T. Okuda, K. Hono, S. V. Eremeev & E. V. Chulkov. Fabrication of a novel magnetic topological heterostructure and temperature evolution of its massive Dirac cone. *Nat Commun* **11**, 4821, doi:10.1038/s41467-020-18645-9 (2020).

[91] Jiazhen Wu, Fucai Liu, Masato Sasase, Koichiro Ienaga, Yukiko Obata, Ryu Yukawa, Koji Horiba, Hiroshi Kumigashira, Satoshi Okuma, Takeshi Inoshita & Hideo Hosono. Natural van der Waals heterostructural single crystals with both magnetic and topological properties. *Science Advances* **5**, eaax9989, doi:10.1126/sciadv.aax9989 (2019).

[92] Raphael C. Vidal, Alexander Zeugner, Jorge I. Facio, Rajyavardhan Ray, M. Hossein Haghighi, Anja U. B Wolter, Laura T. Corredor Bohorquez, Federico Caglieris, Simon Moser, Tim Figgemeier, Thiago R. F Peixoto, Hari Babu Vasili, Manuel Valvidares, Sungwon Jung, Cephise Cacho, Alexey Alfonsov, Kavita Mehlawat, Vladislav Kataev, Christian Hess, Manuel Richter, Bernd Büchner, Jeroen van den Brink, Michael Ruck, Friedrich Reinert, Hendrik Bentmann & Anna Isaeva. Topological Electronic Structure and Intrinsic Magnetization in MnBi4Te7: A Bi2Te3 Derivative with a Periodic Mn Sublattice. *Physical Review X* **9**, 041065, doi:10.1103/PhysRevX.9.041065 (2019).

[93] Yujun Deng, Yijun Yu, Meng Zhu Shi, Zhongxun Guo, Zihan Xu, Jing Wang, Xian Hui Chen & Yuanbo Zhang. Quantum anomalous Hall effect in intrinsic magnetic topological

insulator MnBi2Te4. *Science* **367**, 895-900, doi:10.1126/science.aax8156 (2020).

[94] Jun Ge, Yanzhao Liu, Jiaheng Li, Hao Li, Tianchuang Luo, Yang Wu, Yong Xu & Jian Wang. High-Chern-number and high-temperature quantum Hall effect without Landau levels. *National Science Review* **7**, 1280-1287, doi:10.1093/nsr/nwaa089 (2020).

[95] A. Gao, Y. F. Liu, C. Hu, J. X. Qiu, C. Tzschaschel, B. Ghosh, S. C. Ho, D. Berube, R. Chen, H. Sun, Z. Zhang, X. Y. Zhang, Y. X. Wang, N. Wang, Z. Huang, C. Felser, A. Agarwal, T. Ding, H. J. Tien, A. Akey, J. Gardener, B. Singh, K. Watanabe, T. Taniguchi, K. S. Burch, D. C. Bell, B. B. Zhou, W. Gao, H. Z. Lu, A. Bansil, H. Lin, T. R. Chang, L. Fu, Q. Ma, N. Ni & S. Y. Xu. Layer Hall effect in a 2D topological axion antiferromagnet. *Nature* **595**, 521-525, doi:10.1038/s41586-021-03679-w (2021).

[96] Lixuan Xu, Yuanhao Mao, Hongyuan Wang, Jiaheng Li, Yujie Chen, Yunyouyou Xia, Yiwei Li, Ding Pei, Jing Zhang, Huijun Zheng, Kui Huang, Chaofan Zhang, Shengtao Cui, Aiji Liang, Wei Xia, Hao Su, Sungwon Jung, Cephise Cacho, Meixiao Wang, Gang Li, Yong Xu, Yanfeng Guo, Lexian Yang, Zhongkai Liu, Yulin Chen & Mianheng Jiang. Persistent surface states with diminishing gap in MnBi2Te4/Bi2Te3 superlattice antiferromagnetic topological insulator. *Science Bulletin* **65**, 2086-2093, doi:10.1016/j.scib.2020.07.032 (2020).

[97] S. V. Eremeev, I. P. Rusinov, Y. M. Koroteev, A. Y. Vyazovskaya, M. Hoffmann, P. M. Echenique, A. Ernst, M. M. Otrokov & E. V. Chulkov. Topological Magnetic Materials of the (MnSb(2)Te(4)).(Sb(2)Te(3))(n) van der Waals Compounds Family. *J Phys Chem Lett* **12**, 4268-4277, doi:10.1021/acs.jpclett.1c00875 (2021).

[98] S. V. Eremeev, M. M. Otrokov & E. V. Chulkov. Competing rhombohedral and monoclinic crystal structures in MnPn2Ch4 compounds: An ab-initio study. *Journal of Alloys and Compounds* **709**, 172-178, doi:10.1016/j.jallcom.2017.03.121 (2017).

[99] T. Murakami, Y. Nambu, T. Koretsune, G. Xiangyu, T. Yamamoto, C. M. Brown & H. Kageyama. Realization of interlayer ferromagnetic interaction in MnSb2Te4 toward the magnetic Weyl semimetal state. *Phys Rev B* **100**, doi:10.1103/PhysRevB.100.195103 (2019).

[100] J. Q. Yan, S. Okamoto, M. A. McGuire, A. F. May, R. J. McQueeney & B. C. Sales. Evolution of structural, magnetic, and transport properties in MnBi2−xSbxTe4. *Physical Review B* **100**, doi:10.1103/PhysRevB.100.104409 (2019).

[101] Li Chen, Dongchao Wang, Changmin Shi, Chuan Jiang, Hongmei Liu, Guangliang Cui, Xiaoming Zhang & Xiaolong Li. Electronic structure and magnetism of MnSb2Te4. *Journal of Materials Science* **55**, 14292-14300, doi:10.1007/s10853-020-05005-7 (2020).

[102] Yangyang Chen, Ya-Wen Chuang, Seng Huat Lee, Yanglin Zhu, Kevin Honz, Yingdong Guan, Yu Wang, Ke Wang, Zhiqiang Mao, Jun Zhu, Colin Heikes, P. Quarterman, Pawel Zajdel, Julie A. Borchers & William Ratcliff. Ferromagnetism in van der Waals compound MnSb1.8Bi0.2Te4. *Physical Review Materials* **4**, doi:10.1103/PhysRevMaterials.4.064411 (2020).

[103] Gang Shi, Mingjie Zhang, Dayu Yan, Honglei Feng, Meng Yang, Youguo Shi & Yongqing Li. Anomalous Hall Effect in Layered Ferrimagnet MnSb2Te4*. *Chinese Physics Letters* **37**, doi:10.1088/0256-307x/37/4/047301 (2020).

[104] S. Wimmer, J. Sanchez-Barriga, P. Kuppers, A. Ney, E. Schierle, F. Freyse, O. Caha, J. Michalicka, M. Liebmann, D. Primetzhofer, M. Hoffman, A. Ernst, M. M. Otrokov, G. Bihlmayer, E. Weschke, B. Lake, E. V. Chulkov, M. Morgenstern, G. Bauer, G. Springholz &

O. Rader. Mn-Rich MnSb2 Te4 : A Topological Insulator with Magnetic Gap Closing at High Curie Temperatures of 45-50 K. *Adv Mater* **33**, e2102935, doi:10.1002/adma.202102935 (2021).

[105] Z. Zang, Y. Zhu, M. Xi, S. Tian, T. Wang, P. Gu, Y. Peng, S. Yang, X. Xu, Y. Li, B. Han, L. Liu, Y. Wang, P. Gao, J. Yang, H. Lei, Y. Huang & Y. Ye. Layer-Number-Dependent Antiferromagnetic and Ferromagnetic Behavior in MnSb_2Te_4. *Phys Rev Lett* **128**, 017201, doi:10.1103/PhysRevLett.128.017201 (2022).

[106] S. Huan, S. Zhang, Z. Jiang, H. Su, H. Wang, X. Zhang, Y. Yang, Z. Liu, X. Wang, N. Yu, Z. Zou, D. Shen, J. Liu & Y. Guo. Multiple Magnetic Topological Phases in Bulk van der Waals Crystal MnSb4Te7. *Phys Rev Lett* **126**, 246601, doi:10.1103/PhysRevLett.126.246601 (2021).

[107] Yunyu Yin, Xiaoli Ma, Dayu Yan, Changjiang Yi, Binbin Yue, Jianhong Dai, Lin Zhao, Xiaohui Yu, Youguo Shi, Jian-Tao Wang & Fang Hong. Pressure-driven electronic and structural phase transition in intrinsic magnetic topological insulator
MnSb2Te4. *Physical Review B* **104**, doi:10.1103/PhysRevB.104.174114 (2021).

[108] Jia-Yi Lin, Zhong-Jia Chen, Wen-Qiang Xie, Xiao-Bao Yang & Yu-Jun Zhao. Toward ferromagnetic semimetal ground state with multiple Weyl nodes in van der Waals crystal MnSb4Te7. *New Journal of Physics* **24**, 043033, doi:10.1088/1367-2630/ac6231 (2022).

[109] Cuiying Pei, Ming Xi, Qi Wang, Wujun Shi, Juefei Wu, Lingling Gao, Yi Zhao, Shangjie Tian, Weizheng Cao, Changhua Li, Mingxin Zhang, Shihao Zhu, Yulin Chen, Hechang Lei & Yanpeng Qi. Pressure-induced superconductivity in magnetic topological insulator candidate MnSb4Te7. *Physical Review Materials* **6**, L101801, doi:10.1103/PhysRevMaterials.6.L101801 (2022).

[110] Xin Zhang. Tunable intrinsic ferromagnetic topological phases in bulk van der Waals crystal MnSb6Te10. *arXiv:2111.04973* (2021).

[111] Xiao-Ming Ma, Yufei Zhao, Ke Zhang, Shiv Kumar, Ruie Lu, Jiayu Li, Qiushi Yao, Jifeng Shao, Fuchen Hou, Xuefeng Wu, Meng Zeng, Yu-Jie Hao, Zhanyang Hao, Yuan Wang, Xiang-Rui Liu, Huiwen Shen, Hongyi Sun, Jiawei Mei, Koji Miyamoto, Taichi Okuda, Masashi Arita, Eike F. Schwier, Kenya Shimada, Ke Deng, Cai Liu, Junhao Lin, Yue Zhao, Chaoyu Chen, Qihang Liu & Chang Liu. Realization of a tunable surface Dirac gap in Sb-doped MnBi2Te4. *Physical Review B* **103**, L121112, doi:10.1103/PhysRevB.103.L121112 (2021).

[112] T. Zhu, A. J. Bishop, T. Zhou, M. Zhu, D. J. O'Hara, A. A. Baker, S. Cheng, R. C. Walko, J. J. Repicky, T. Liu, J. A. Gupta, C. M. Jozwiak, E. Rotenberg, J. Hwang, I. Zutic & R. K. Kawakami. Synthesis, Magnetic Properties, and Electronic Structure of Magnetic Topological Insulator MnBi2Se4. *Nano Lett*, doi:10.1021/acs.nanolett.1c00141 (2021).

[113] M. Q. Arguilla, N. D. Cultrara, Z. J. Baum, S. Jiang, R. D. Ross & J. E. Goldberger. EuSn2As2: an exfoliatable magnetic layered Zintl–Klemm phase. *Inorganic Chemistry Frontiers* **4**, 378-386, doi:10.1039/c6qi00476h (2017).

[114] Firoza Kabir. Observation of multiple Dirac states in a magnetic topological material EuMg2Bi2. *arXiv:1912.08645* (2019).

[115] Sabin Regmi, M. Mofazzel Hosen, Barun Ghosh, Bahadur Singh, Gyanendra Dhakal, Christopher Sims, Baokai Wang, Firoza Kabir, Klauss Dimitri, Yangyang Liu, Amit Agarwal, Hsin Lin, Dariusz Kaczorowski, Arun Bansil & Madhab Neupane. Temperature-dependent

[115] electronic structure in a higher-order topological insulator candidate EuIn2As2. *Physical Review B* **102**, doi:10.1103/PhysRevB.102.165153 (2020).

[116] Madalynn Marshall, Ivo Pletikosić, Mohammad Yahyavi, Hung-Ju Tien, Tay-Rong Chang, Huibo Cao & Weiwei Xie. Magnetic and electronic structures of antiferromagnetic topological material candidate EuMg2Bi2. *Journal of Applied Physics* **129**, doi:10.1063/5.0035703 (2021).

[117] Yang Zhang, Ke Deng, Xiao Zhang, Meng Wang, Yuan Wang, Cai Liu, Jia-Wei Mei, Shiv Kumar, Eike F. Schwier, Kenya Shimada, Chaoyu Chen & Bing Shen. In-plane antiferromagnetic moments and magnetic polaron in the axion topological insulator candidate EuIn2As2. *Physical Review B* **101**, doi:10.1103/PhysRevB.101.205126 (2020).

[118] L. Zhao, C. Yi, C. T. Wang, Z. Chi, Y. Yin, X. Ma, J. Dai, P. Yang, B. Yue, J. Cheng, F. Hong, J. T. Wang, Y. Han, Y. Shi & X. Yu. Monoclinic EuSn2As2: A Novel High-Pressure Network Structure. *Phys Rev Lett* **126**, 155701, doi:10.1103/PhysRevLett.126.155701 (2021).

[119] S. X. M. Riberolles, T. V. Trevisan, B. Kuthanazhi, T. W. Heitmann, F. Ye, D. C. Johnston, S. L. Bud'ko, D. H. Ryan, P. C. Canfield, A. Kreyssig, A. Vishwanath, R. J. McQueeney, L. Wang, P. P. Orth & B. G. Ueland. Magnetic crystalline-symmetry-protected axion electrodynamics and field-tunable unpinned Dirac cones in EuIn2As2. *Nat Commun* **12**, 999, doi:10.1038/s41467-021-21154-y (2021).

[120] Huan-Cheng Chen, Zhe-Feng Lou, Yu-Xing Zhou, Qin Chen, Bin-Jie Xu, Shui-Jin Chen, Jian-Hua Du, Jin-Hu Yang, Hang-Dong Wang & Ming-Hu Fang. Negative Magnetoresistance in Antiferromagnetic Topological Insulator EuSn2As2

*. *Chinese Physics Letters* **37**, doi:10.1088/0256-307x/37/4/047201 (2020).

[121] Huijie Li, Wenshuai Gao, Zheng Chen, Weiwei Chu, Yong Nie, Shuaiqi Ma, Yuyan Han, Min Wu, Tian Li, Qun Niu, Wei Ning, Xiangde Zhu & Mingliang Tian. Magnetic properties of the layered magnetic topological insulator EuSn2As2. *Physical Review B* **104**, doi:10.1103/PhysRevB.104.054435 (2021).

[122] Hualei Sun, Cuiqun Chen, Yusheng Hou, Weiliang Wang, Yu Gong, Mengwu Huo, Lisi Li, Jia Yu, Wanping Cai, Naitian Liu, Ruqian Wu, Dao-Xin Yao & Meng Wang. Magnetism variation of the compressed antiferromagnetic topological insulator EuSn2As2. *Science China Physics, Mechanics & Astronomy* **64**, doi:10.1007/s11433-021-1760-x (2021).

[123] Andrea M. Goforth, Peter Klavins, James C. Fettinger & Susan M. Kauzlarich. Magnetic Properties and Negative Colossal Magnetoresistance of the Rare Earth Zintl phase EuIn2As2. *Inorganic Chemistry* **47**, 11048-11056, doi:10.1021/ic801290u (2008).

[124] Tomasz Tolinski and Dariusz Kaczorowski. Magnetic properties of the putative higher-order topological insulator EuIn2As2. *SciPost Physics Proceedings*, doi:10.21468/SciPostPhysProc (2022).

[125] Y. Xu, Z. Song, Z. Wang, H. Weng & X. Dai. Higher-Order Topology of the Axion Insulator EuIn2As2. *Phys Rev Lett* **122**, 256402, doi:10.1103/PhysRevLett.122.256402 (2019).

[126] Mingda Gong, Divyanshi Sar, Joel Friedman, Dariusz Kaczorowski, S. Abdel Razek, Wei-Cheng Lee & Pegor Aynajian. Surface state evolution induced by magnetic order in axion insulator candidate EuIn2As2. *Physical Review B* **106**, 125156, doi:10.1103/PhysRevB.106.125156 (2022).

[127] Priscila Rosa, Yuanfeng Xu, Marein Rahn, Jean Souza, Satya Kushwaha, Larissa Veiga, Alessandro Bombardi, Sean Thomas, Marc Janoschek, Eric Bauer, Mun Chan, Zhijun Wang,

[127] Joe Thompson, Neil Harrison, Pascoal Pagliuso, Andrei Bernevig & Filip Ronning. Colossal magnetoresistance in a nonsymmorphic antiferromagnetic insulator. *npj Quantum Materials* **5**, doi:10.1038/s41535-020-00256-8 (2020).

[128] Nicodemos Varnava, Tanya Berry, Tyrel M. McQueen & David Vanderbilt. Engineering magnetic topological insulators in Eu5M2X6 Zintl compounds. *Physical Review B* **105**, doi:10.1103/PhysRevB.105.235128 (2022).

[129] H. Wang, N. Mao, X. Hu, Y. Dai, B. Huang & C. Niu. A magnetic topological insulator in two-dimensional EuCd(2)Bi(2): giant gap with robust topology against magnetic transitions. *Mater Horiz* **8**, 956-961, doi:10.1039/d0mh01214a (2021).

[130] J. Liu, S. Meng & J. T. Sun. Spin-Orientation-Dependent Topological States in Two-Dimensional Antiferromagnetic NiTl(2)S(4) Monolayers. *Nano Lett* **19**, 3321-3326, doi:10.1021/acs.nanolett.9b00948 (2019).

[131] Peizhe Tang, Quan Zhou, Gang Xu & Shou-Cheng Zhang. Dirac fermions in an antiferromagnetic semimetal. *Nature Physics* **12**, 1100-1104, doi:10.1038/nphys3839 (2016).

[132] Jing Wang. Antiferromagnetic Dirac semimetals in two dimensions. *Physical Review B* **95**, doi:10.1103/PhysRevB.95.115138 (2017).

[133] S. M. Young & B. J. Wieder. Filling-Enforced Magnetic Dirac Semimetals in Two Dimensions. *Phys Rev Lett* **118**, 186401, doi:10.1103/PhysRevLett.118.186401 (2017).

[134] Si Li, Ying Liu, Zhi-Ming Yu, Yalong Jiao, Shan Guan, Xian-Lei Sheng, Yugui Yao & Shengyuan A. Yang. Two-dimensional antiferromagnetic Dirac fermions in monolayer TaCoTe2. *Physical Review B* **100**, doi:10.1103/PhysRevB.100.205102 (2019).

[135] Noam Morali, Rajib Batabyal, Pranab Kumar Nag, Enke Liu, Qiunan Xu, Yan Sun, Binghai Yan, Claudia Felser, Nurit Avraham & Haim Beidenkopf. Fermi-arc diversity on surface terminations of the magnetic Weyl semimetal Co3Sn2S2. *Science* **365**, 1286, doi:10.1126/science.aav2334 (2019).

[136] D. F. Liu, A. J. Liang, E. K. Liu, Q. N. Xu, Y. W. Li, C. Chen, D. Pei, W. J. Shi, S. K. Mo, P. Dudin, T. Kim, C. Cacho, G. Li, Y. Sun, L. X. Yang, Z. K. Liu, S. S. P. Parkin, C. Felser & Y. L. Chen. Magnetic Weyl semimetal phase in a Kagomé crystal. *Science* **365**, 1282, doi:10.1126/science.aav2873 (2019).

[137] K. Kuroda, T. Tomita, M. T. Suzuki, C. Bareille, A. A. Nugroho, P. Goswami, M. Ochi, M. Ikhlas, M. Nakayama, S. Akebi, R. Noguchi, R. Ishii, N. Inami, K. Ono, H. Kumigashira, A. Varykhalov, T. Muro, T. Koretsune, R. Arita, S. Shin, T. Kondo & S. Nakatsuji. Evidence for magnetic Weyl fermions in a correlated metal. *Nat Mater* **16**, 1090-1095, doi:10.1038/nmat4987 (2017).

[138] Ajaya K. Nayak, Julia Erika Fischer, Yan Sun, Binghai Yan, Julie Karel, Alexander C. Komarek, Chandra Shekhar, Nitesh Kumar, Walter Schnelle, Jürgen Kübler, Claudia Felser & Stuart S. P. Parkin. Large anomalous Hall effect driven by a nonvanishing Berry curvature in the noncolinear antiferromagnet Mn3Ge. *Science Advances* **2**, e1501870, doi:10.1126/sciadv.1501870 (2016).

[139] S. Nakatsuji, N. Kiyohara & T. Higo. Large anomalous Hall effect in a non-collinear antiferromagnet at room temperature. *Nature* **527**, 212-215, doi:10.1038/nature15723 (2015).

[140] B. Q. Lv, N. Xu, H. M. Weng, J. Z. Ma, P. Richard, X. C. Huang, L. X. Zhao, G. F. Chen, C. E.

Matt, F. Bisti, V. N. Strocov, J. Mesot, Z. Fang, X. Dai, T. Qian, M. Shi & H. Ding. Observation of Weyl nodes in TaAs. *Nature Physics* **11**, 724-727, doi:10.1038/nphys3426 (2015).

[141] J. Z. Ma, J. B. He, Y. F. Xu, B. Q. Lv, D. Chen, W. L. Zhu, S. Zhang, L. Y. Kong, X. Gao, L. Y. Rong, Y. B. Huang, P. Richard, C. Y. Xi, E. S. Choi, Y. Shao, Y. L. Wang, H. J. Gao, X. Dai, C. Fang, H. M. Weng, G. F. Chen, T. Qian & H. Ding. Three-component fermions with surface Fermi arcs in tungsten carbide. *Nature Physics* **14**, 349-354, doi:10.1038/s41567-017-0021-8 (2018).

[142] Su-Yang Xu, Nasser Alidoust, Ilya Belopolski, Zhujun Yuan, Guang Bian, Tay-Rong Chang, Hao Zheng, Vladimir N. Strocov, Daniel S. Sanchez, Guoqing Chang, Chenglong Zhang, Daixiang Mou, Yun Wu, Lunan Huang, Chi-Cheng Lee, Shin-Ming Huang, BaoKai Wang, Arun Bansil, Horng-Tay Jeng, Titus Neupert, Adam Kaminski, Hsin Lin, Shuang Jia & M. Zahid Hasan. Discovery of a Weyl fermion state with Fermi arcs in niobium arsenide. *Nature Physics* **11**, 748-754, doi:10.1038/nphys3437 (2015).

[143] L. X. Yang, Z. K. Liu, Y. Sun, H. Peng, H. F. Yang, T. Zhang, B. Zhou, Y. Zhang, Y. F. Guo, M. Rahn, D. Prabhakaran, Z. Hussain, S. K. Mo, C. Felser, B. Yan & Y. L. Chen. Weyl semimetal phase in the non-centrosymmetric compound TaAs. *Nature Physics* **11**, 728-732, doi:10.1038/nphys3425 (2015).

[144] Z. K. Liu, B. Zhou, Y. Zhang, Z. J. Wang, H. M. Weng, D. Prabhakaran, S. K. Mo, Z. X. Shen, Z. Fang, X. Dai, Z. Hussain & Y. L. Chen. Discovery of a Three-Dimensional Topological Dirac Semimetal, Na3Bi. *Science* **343**, 864 (2014).

[145] Z. K. Liu, J. Jiang, B. Zhou, Z. J. Wang, Y. Zhang, H. M. Weng, D. Prabhakaran, S. K. Mo, H. Peng, P. Dudin, T. Kim, M. Hoesch, Z. Fang, X. Dai, Z. X. Shen, D. L. Feng, Z. Hussain & Y. L. Chen. A stable three-dimensional topological Dirac semimetal Cd3As2. *Nature Materials* **13**, 677-681, doi:10.1038/nmat3990 (2014).

[146] M. Neupane, S. Y. Xu, R. Sankar, N. Alidoust, G. Bian, C. Liu, I. Belopolski, T. R. Chang, H. T. Jeng, H. Lin, A. Bansil, F. Chou & M. Z. Hasan. Observation of a three-dimensional topological Dirac semimetal phase in high-mobility Cd3As2. *Nat Commun* **5**, 3786, doi:10.1038/ncomms4786 (2014).

[147] Gang Xu, Hongming Weng, Zhijun Wang, Xi Dai & Zhong Fang. Chern Semimetal and the Quantized Anomalous Hall Effect in HgCr_{2}Se_{4}. *Physical Review Letters* **107**, 186806 (2011).

[148] Seung-Hwan Do, Koji Kaneko, Ryoichi Kajimoto, Kazuya Kamazawa, Matthew B. Stone, Jiao Y. Y. Lin, Shinichi Itoh, Takatsugu Masuda, German D. Samolyuk, Elbio Dagotto, William R. Meier, Brian C. Sales, Hu Miao & Andrew D. Christianson. Damped Dirac magnon in the metallic kagome antiferromagnet FeSn. *Physical Review B* **105**, doi:10.1103/PhysRevB.105.L180403 (2022).

[149] Zhiyong Lin, Chongze Wang, Pengdong Wang, Seho Yi, Lin Li, Qiang Zhang, Yifan Wang, Zhongyi Wang, Hao Huang, Yan Sun, Yaobo Huang, Dawei Shen, Donglai Feng, Zhe Sun, Jun-Hyung Cho, Changgan Zeng & Zhenyu Zhang. Dirac fermions in antiferromagnetic FeSn kagome lattices with combined space inversion and time-reversal symmetry. *Physical Review B* **102**, doi:10.1103/PhysRevB.102.155103 (2020).

[150] M. Kang, L. Ye, S. Fang, J. S. You, A. Levitan, M. Han, J. I. Facio, C. Jozwiak, A. Bostwick, E. Rotenberg, M. K. Chan, R. D. McDonald, D. Graf, K. Kaznatcheev, E. Vescovo, D. C. Bell, E. Kaxiras, J. van den Brink, M. Richter, M. Prasad Ghimire, J. G. Checkelsky & R. Comin. Dirac

fermions and flat bands in the ideal kagome metal FeSn. *Nat Mater* **19**, 163-169, doi:10.1038/s41563-019-0531-0 (2020).

[151] M. Han, H. Inoue, S. Fang, C. John, L. Ye, M. K. Chan, D. Graf, T. Suzuki, M. P. Ghimire, W. J. Cho, E. Kaxiras & J. G. Checkelsky. Evidence of two-dimensional flat band at the surface of antiferromagnetic kagome metal FeSn. *Nat Commun* **12**, 5345, doi:10.1038/s41467-021-25705-1 (2021).

[152] Si-Hong Lee, Youngjae Kim, Beopgil Cho, Jaemun Park, Min-Seok Kim, Kidong Park, Hoyeon Jeon, Minkyung Jung, Keeseong Park, JaeDong Lee & Jungpil Seo. Spin-polarized and possible pseudospin-polarized scanning tunneling microscopy in kagome metal FeSn. *Communications Physics* **5**, doi:10.1038/s42005-022-01012-z (2022).

[153] Brian C. Sales, Jiaqiang Yan, William R. Meier, Andrew D. Christianson, Satoshi Okamoto & Michael A. McGuire. Electronic, magnetic, and thermodynamic properties of the kagome layer compound FeSn. *Physical Review Materials* **3**, doi:10.1103/PhysRevMaterials.3.114203 (2019).

[154] Chang Liu, ChangJiang Yi, XingYu Wang, JianLei Shen, Tao Xie, Lin Yang, Tom Fennel, Uwe Stuhr, ShiLiang Li, HongMing Weng, YouGuo Shi, EnKe Liu & HuiQian Luo. Anisotropic magnetoelastic response in the magnetic Weyl semimetal $Co_3Sn_2S_2$. *Science China Physics, Mechanics & Astronomy* **64**, doi:10.1007/s11433-020-1655-2 (2021).

[155] D. F. Liu, E. K. Liu, Q. N. Xu, J. L. Shen, Y. W. Li, D. Pei, A. J. Liang, P. Dudin, T. K. Kim, C. Cacho, Y. F. Xu, Y. Sun, L. X. Yang, Z. K. Liu, C. Felser, S. S. P. Parkin & Y. L. Chen. Direct observation of the spin–orbit coupling effect in magnetic Weyl semimetal $Co_3Sn_2S_2$. *npj Quantum Materials* **7**, doi:10.1038/s41535-021-00392-9 (2022).

[156] M. Kanagaraj, Jiai Ning & Liang He. Topological $Co_3Sn_2S_2$ magnetic Weyl semimetal: From fundamental understanding to diverse fields of study. *Reviews in Physics* **8**, doi:10.1016/j.revip.2022.100072 (2022).

[157] Ilya Belopolski, Tyler A. Cochran, Xiaoxiong Liu, Zi-Jia Cheng, Xian P. Yang, Zurab Guguchia, Stepan S. Tsirkin, Jia-Xin Yin, Praveen Vir, Gohil S. Thakur, Songtian S. Zhang, Junyi Zhang, Konstantine Kaznatcheev, Guangming Cheng, Guoqing Chang, Daniel Multer, Nana Shumiya, Maksim Litskevich, Elio Vescovo, Timur K. Kim, Cephise Cacho, Nan Yao, Claudia Felser, Titus Neupert & M. Zahid Hasan. Signatures of Weyl Fermion Annihilation in a Correlated Kagome Magnet. *Physical Review Letters* **127**, 256403, doi:10.1103/PhysRevLett.127.256403 (2021).

[158] Guowei Li, Qiunan Xu, Wujun Shi, Chenguang Fu, Lin Jiao, Machteld E. Kamminga, Mingquan Yu, Harun Tüysüz, Nitesh Kumar, Vicky Süß, Rana Saha, Abhay K. Srivastava, Steffen Wirth, Gudrun Auffermann, Johannes Gooth, Stuart Parkin, Yan Sun, Enke Liu & Claudia Felser. Surface states in bulk single crystal of topological semimetal $Co_3Sn_2S_2$ toward water oxidation. *Science Advances* **5**, eaaw9867, doi:10.1126/sciadv.aaw9867 (2019).

[159] Qiunan Xu, Enke Liu, Wujun Shi, Lukas Muechler, Jacob Gayles, Claudia Felser & Yan Sun. Topological surface Fermi arcs in the magnetic Weyl semimetal $Co_3Sn_2S_2$. *Physical Review B* **97**, doi:10.1103/PhysRevB.97.235416 (2018).

[160] Q. Wang, Y. Xu, R. Lou, Z. Liu, M. Li, Y. Huang, D. Shen, H. Weng, S. Wang & H. Lei. Large intrinsic anomalous Hall effect in half-metallic ferromagnet $Co_3Sn_2S_2$ with magnetic Weyl fermions. *Nat Commun* **9**, 3681, doi:10.1038/s41467-018-06088-2 (2018).

[161] M. Tanaka, Y. Fujishiro, M. Mogi, Y. Kaneko, T. Yokosawa, N. Kanazawa, S. Minami, T. Koretsune, R. Arita, S. Tarucha, M. Yamamoto & Y. Tokura. Topological Kagome Magnet Co(3)Sn(2)S(2) Thin Flakes with High Electron Mobility and Large Anomalous Hall Effect. *Nano Lett* **20**, 7476-7481, doi:10.1021/acs.nanolett.0c02962 (2020).

[162] H. Reichlova, T. Janda, J. Godinho, A. Markou, D. Kriegner, R. Schlitz, J. Zelezny, Z. Soban, M. Bejarano, H. Schultheiss, P. Nemec, T. Jungwirth, C. Felser, J. Wunderlich & S. T. B. Goennenwein. Imaging and writing magnetic domains in the non-collinear antiferromagnet Mn3Sn. *Nat Commun* **10**, 5459, doi:10.1038/s41467-019-13391-z (2019).

[163] T. Chen, T. Tomita, S. Minami, M. Fu, T. Koretsune, M. Kitatani, I. Muhammad, D. Nishio-Hamane, R. Ishii, F. Ishii, R. Arita & S. Nakatsuji. Anomalous transport due to Weyl fermions in the chiral antiferromagnets Mn(3)X, X = Sn, Ge. *Nat Commun* **12**, 572, doi:10.1038/s41467-020-20838-1 (2021).

[164] J. R. Soh, F. de Juan, N. Qureshi, H. Jacobsen, H. Y. Wang, Y. F. Guo & A. T. Boothroyd. Ground-state magnetic structure of Mn3Ge. *Physical Review B* **101**, doi:10.1103/PhysRevB.101.140411 (2020).

[165] J. Liu & L. Balents. Anomalous Hall Effect and Topological Defects in Antiferromagnetic Weyl Semimetals: Mn3Sn/Ge. *Phys Rev Lett* **119**, 087202, doi:10.1103/PhysRevLett.119.087202 (2017).

[166] Hao Yang, Yan Sun, Yang Zhang, Wu-Jun Shi, Stuart S. P. Parkin & Binghai Yan. Topological Weyl semimetals in the chiral antiferromagnetic materials Mn3Ge and Mn3Sn. *New Journal of Physics* **19**, doi:10.1088/1367-2630/aa5487 (2017).

[167] Naoki Kiyohara, Takahiro Tomita & Satoru Nakatsuji. Giant Anomalous Hall Effect in the Chiral Antiferromagnet
Mn3Ge. *Physical Review Applied* **5**, doi:10.1103/PhysRevApplied.5.064009 (2016).

[168] Tomoya Higo, Danru Qu, Yufan Li, C. L. Chien, Yoshichika Otani & Satoru Nakatsuji. Anomalous Hall effect in thin films of the Weyl antiferromagnet Mn3Sn. *Applied Physics Letters* **113**, doi:10.1063/1.5064697 (2018).

[169] T. Matsuda, N. Kanda, T. Higo, N. P. Armitage, S. Nakatsuji & R. Matsunaga. Room-temperature terahertz anomalous Hall effect in Weyl antiferromagnet Mn(3)Sn thin films. *Nat Commun* **11**, 909, doi:10.1038/s41467-020-14690-6 (2020).

[170] James M. Taylor, Anastasios Markou, Edouard Lesne, Pranava Keerthi Sivakumar, Chen Luo, Florin Radu, Peter Werner, Claudia Felser & Stuart S. P. Parkin. Anomalous and topological Hall effects in epitaxial thin films of the noncollinear antiferromagnet Mn3Sn. *Physical Review B* **101**, doi:10.1103/PhysRevB.101.094404 (2020).

[171] Muhammad Ikhlas, Takahiro Tomita, Takashi Koretsune, Michi-To Suzuki, Daisuke Nishio-Hamane, Ryotaro Arita, Yoshichika Otani & Satoru Nakatsuji. Large anomalous Nernst effect at room temperature in a chiral antiferromagnet. *Nature Physics* **13**, 1085-1090, doi:10.1038/nphys4181 (2017).

[172] Christoph Wuttke, Federico Caglieris, Steffen Sykora, Francesco Scaravaggi, Anja U. B. Wolter, Kaustuv Manna, Vicky Süss, Chandra Shekhar, Claudia Felser, Bernd Büchner & Christian Hess. Berry curvature unravelled by the anomalous Nernst effect in Mn3Ge. *Physical Review B* **100**, doi:10.1103/PhysRevB.100.085111 (2019).

[173] X. Li, C. Collignon, L. Xu, H. Zuo, A. Cavanna, U. Gennser, D. Mailly, B. Fauque, L. Balents,

Z. Zhu & K. Behnia. Chiral domain walls of Mn3Sn and their memory. *Nat Commun* **10**, 3021, doi:10.1038/s41467-019-10815-8 (2019).

[174] Pradeep K. Rout, P. V. Prakash Madduri, Subhendu K. Manna & Ajaya K. Nayak. Field-induced topological Hall effect in the noncoplanar triangular antiferromagnetic geometry of Mn3Sn. *Physical Review B* **99**, doi:10.1103/PhysRevB.99.094430 (2019).

[175] Liangcai Xu, Xiaokang Li, Linchao Ding, Kamran Behnia & Zengwei Zhu. Planar Hall effect caused by the memory of antiferromagnetic domain walls in Mn3Ge. *Applied Physics Letters* **117**, doi:10.1063/5.0030546 (2020).

[176] M. Kimata, H. Chen, K. Kondou, S. Sugimoto, P. K. Muduli, M. Ikhlas, Y. Omori, T. Tomita, A. H. MacDonald, S. Nakatsuji & Y. Otani. Magnetic and magnetic inverse spin Hall effects in a non-collinear antiferromagnet. *Nature* **565**, 627-630, doi:10.1038/s41586-018-0853-0 (2019).

[177] P. Li, J. Koo, W. Ning, J. Li, L. Miao, L. Min, Y. Zhu, Y. Wang, N. Alem, C. X. Liu, Z. Mao & B. Yan. Giant room temperature anomalous Hall effect and tunable topology in a ferromagnetic topological semimetal Co(2)MnAl. *Nat Commun* **11**, 3476, doi:10.1038/s41467-020-17174-9 (2020).

[178] G. Chang, S. Y. Xu, X. Zhou, S. M. Huang, B. Singh, B. Wang, I. Belopolski, J. Yin, S. Zhang, A. Bansil, H. Lin & M. Z. Hasan. Topological Hopf and Chain Link Semimetal States and Their Application to Co_2MnGa. *Phys Rev Lett* **119**, 156401, doi:10.1103/PhysRevLett.119.156401 (2017).

[179] I. Belopolski, G. Chang, T. A. Cochran, Z. J. Cheng, X. P. Yang, C. Hugelmeyer, K. Manna, J. X. Yin, G. Cheng, D. Multer, M. Litskevich, N. Shumiya, S. S. Zhang, C. Shekhar, N. B. M. Schroter, A. Chikina, C. Polley, B. Thiagarajan, M. Leandersson, J. Adell, S. M. Huang, N. Yao, V. N. Strocov, C. Felser & M. Z. Hasan. Observation of a linked-loop quantum state in a topological magnet. *Nature* **604**, 647-652, doi:10.1038/s41586-022-04512-8 (2022).

[180] Z. Wang, M. G. Vergniory, S. Kushwaha, M. Hirschberger, E. V. Chulkov, A. Ernst, N. P. Ong, R. J. Cava & B. A. Bernevig. Time-Reversal-Breaking Weyl Fermions in Magnetic Heusler Alloys. *Phys Rev Lett* **117**, 236401, doi:10.1103/PhysRevLett.117.236401 (2016).

[181] G. Chang, S. Y. Xu, H. Zheng, B. Singh, C. H. Hsu, G. Bian, N. Alidoust, I. Belopolski, D. S. Sanchez, S. Zhang, H. Lin & M. Z. Hasan. Room-temperature magnetic topological Weyl fermion and nodal line semimetal states in half-metallic Heusler Co2TiX (X=Si, Ge, or Sn). *Sci Rep* **6**, 38839, doi:10.1038/srep38839 (2016).

[182] R. Y. Umetsu, K. Kobayashi, A. Fujita, R. Kainuma & K. Ishida. Magnetic properties and stability of L21 and B2 phases in the Co2MnAl Heusler alloy. *Journal of Applied Physics* **103**, doi:10.1063/1.2836677 (2008).

[183] A. W. Carbonari, R. N. Saxena, W. Pendl, J. Mestnik Filho, R. N. Attili, M. Olzon-Dionysio & S. D. de Souza. Magnetic hyperfine field in the Heusler alloys Co2YZ (Y = V, Nb, Ta, Cr; Z = Al, Ga). *Journal of Magnetism and Magnetic Materials* **163**, 313-321, doi:10.1016/s0304-8853(96)00338-1 (1996).

[184] Zhongbo Yan, Ren Bi, Huitao Shen, Ling Lu, Shou-Cheng Zhang & Zhong Wang. Nodal-link semimetals. *Physical Review B* **96**, doi:10.1103/PhysRevB.96.041103 (2017).

[185] Motohiko Ezawa. Topological semimetals carrying arbitrary Hopf numbers: Fermi surface topologies of a Hopf link, Solomon's knot, trefoil knot, and other linked nodal varieties. *Physical Review B* **96**, doi:10.1103/PhysRevB.96.041202 (2017).

[186] Po-Yao Chang & Chuck-Hou Yee. Weyl-link semimetals. *Physical Review B* **96**, doi:10.1103/PhysRevB.96.081114 (2017).

[187] Kazuki Sumida, Yuya Sakuraba, Keisuke Masuda, Takashi Kono, Masaaki Kakoki, Kazuki Goto, Weinan Zhou, Koji Miyamoto, Yoshio Miura, Taichi Okuda & Akio Kimura. Spin-polarized Weyl cones and giant anomalous Nernst effect in ferromagnetic Heusler films. *Communications Materials* **1**, doi:10.1038/s43246-020-00088-w (2020).

[188] Q. Wu, A. A. Soluyanov & T. Bzdusek. Non-Abelian band topology in noninteracting metals. *Science* **365**, 1273-1277, doi:10.1126/science.aau8740 (2019).

[189] I. Belopolski, K. Manna, D. S. Sanchez, G. Chang, B. Ernst, J. Yin, S. S. Zhang, T. Cochran, N. Shumiya, H. Zheng, B. Singh, G. Bian, D. Multer, M. Litskevich, X. Zhou, S. M. Huang, B. Wang, T. R. Chang, S. Y. Xu, A. Bansil, C. Felser, H. Lin & M. Z. Hasan. Discovery of topological Weyl fermion lines and drumhead surface states in a room temperature magnet. *Science* **365**, 1278-1281, doi:10.1126/science.aav2327 (2019).

[190] C. Zhong, Y. Chen, Z. M. Yu, Y. Xie, H. Wang, S. A. Yang & S. Zhang. Three-dimensional Pentagon Carbon with a genesis of emergent fermions. *Nat Commun* **8**, 15641, doi:10.1038/ncomms15641 (2017).

[191] Jian Yuan, Xianbiao Shi, Hao Su, Xin Zhang, Xia Wang, Na Yu, Zhiqiang Zou, Weiwei Zhao, Jianpeng Liu & Yanfeng Guo. Magnetization tunable Weyl states in $EuB_6$. *Physical Review B* **106**, doi:10.1103/PhysRevB.106.054411 (2022).

[192] Shun-Ye Gao, Sheng Xu, Hang Li, Chang-Jiang Yi, Si-Min Nie, Zhi-Cheng Rao, Huan Wang, Quan-Xin Hu, Xue-Zhi Chen, Wen-Hui Fan, Jie-Rui Huang, Yao-Bo Huang, Nini Pryds, Ming Shi, Zhi-Jun Wang, You-Guo Shi, Tian-Long Xia, Tian Qian & Hong Ding. Time-Reversal Symmetry Breaking Driven Topological Phase Transition in $EuB_6$. *Physical Review X* **11**, doi:10.1103/PhysRevX.11.021016 (2021).

[193] S. Nie, Y. Sun, F. B. Prinz, Z. Wang, H. Weng, Z. Fang & X. Dai. Magnetic Semimetals and Quantized Anomalous Hall Effect in $EuB_6$. *Phys Rev Lett* **124**, 076403, doi:10.1103/PhysRevLett.124.076403 (2020).

[194] X. Zhang, S. von Molnar, Z. Fisk & P. Xiong. Spin-dependent electronic states of the ferromagnetic semimetal $EuB_6$. *Phys Rev Lett* **100**, 167001, doi:10.1103/PhysRevLett.100.167001 (2008).

[195] Jungho Kim, Wei Ku, Chi-Cheng Lee, D. S. Ellis, B. K. Cho, A. H. Said, Y. Shvyd'ko & Young-June Kim. Spin-split conduction band in $EuB_6$ and tuning of half-metallicity with external stimuli. *Physical Review B* **87**, doi:10.1103/PhysRevB.87.155104 (2013).

[196] S. Süllow, I. Prasad, M. C. Aronson, J. L. Sarrao, Z. Fisk, D. Hristova, A. H. Lacerda, M. F. Hundley, A. Vigliante & D. Gibbs. Structure and magnetic order of $EuB_6$. *Physical Review B* **57**, 5860-5869, doi:10.1103/PhysRevB.57.5860 (1998).

[197] M. L. Brooks, T. Lancaster, S. J. Blundell, W. Hayes, F. L. Pratt & Z. Fisk. Magnetic phase separation in $EuB_6$ detected by muon spin rotation. *Physical Review B* **70**, doi:10.1103/PhysRevB.70.020401 (2004).

[198] L. Degiorgi, E. Felder, H. R. Ott, J. L. Sarrao & Z. Fisk. Low-Temperature Anomalies and Ferromagnetism of $EuB_6$. *Physical Review Letters* **79**, 5134-5137, doi:10.1103/PhysRevLett.79.5134 (1997).

[199] C. N. Guy, S. von Molnar, J. Etourneau & Z. Fisk. Charge transport and pressure dependence of Tc of single crystal, ferromagnetic $EuB_6$. *Solid State Communications* **33**,

1055-1058, doi:10.1016/0038-1098(80)90316-6 (1980).

[200] P. Nyhus, S. Yoon, M. Kauffman, S. L. Cooper, Z. Fisk & J. Sarrao. Spectroscopic study of bound magnetic polaron formation and the metal-semiconductor transition inEuB6. *Physical Review B* **56**, 2717-2721, doi:10.1103/PhysRevB.56.2717 (1997).

[201] G. Beaudin, L. M. Fournier, A. D. Bianchi, M. Nicklas, M. Kenzelmann, M. Laver & W. Witczak-Krempa. Possible quantum nematic phase in a colossal magnetoresistance material. *Physical Review B* **105**, doi:10.1103/PhysRevB.105.035104 (2022).

[202] W. L. Liu, X. Zhang, S. M. Nie, Z. T. Liu, X. Y. Sun, H. Y. Wang, J. Y. Ding, Q. Jiang, L. Sun, F. H. Xue, Z. Huang, H. Su, Y. C. Yang, Z. C. Jiang, X. L. Lu, J. Yuan, S. Cho, J. S. Liu, Z. H. Liu, M. Ye, S. L. Zhang, H. M. Weng, Z. Liu, Y. F. Guo, Z. J. Wang & D. W. Shen. Spontaneous Ferromagnetism Induced Topological Transition in EuB_6. *Phys Rev Lett* **129**, 166402, doi:10.1103/PhysRevLett.129.166402 (2022).

[203] Qingqi Zeng, Changjiang Yi, Jianlei Shen, Binbin Wang, Hongxiang Wei, Youguo Shi & Enke Liu. Berry curvature induced antisymmetric in-plane magneto-transport in magnetic Weyl EuB6. *Applied Physics Letters* **121**, doi:10.1063/5.0114252 (2022).

[204] Bin Chen, JinHu Yang, HangDong Wang, Masaki Imai, Hiroto Ohta, Chishiro Michioka, Kazuyoshi Yoshimura & MingHu Fang. Magnetic Properties of Layered Itinerant Electron Ferromagnet Fe3GeTe2. *Journal of the Physical Society of Japan* **82**, doi:10.7566/jpsj.82.124711 (2013).

[205] Y. Zhang, H. Lu, X. Zhu, S. Tan, W. Feng, Q. Liu, W. Zhang, Q. Chen, Y. Liu, X. Luo, D. Xie, L. Luo, Z. Zhang & X. Lai. Emergence of Kondo lattice behavior in a van der Waals itinerant ferromagnet, Fe(3)GeTe(2). *Sci Adv* **4**, eaao6791, doi:10.1126/sciadv.aao6791 (2018).

[206] Y. Deng, Y. Yu, Y. Song, J. Zhang, N. Z. Wang, Z. Sun, Y. Yi, Y. Z. Wu, S. Wu, J. Zhu, J. Wang, X. H. Chen & Y. Zhang. Gate-tunable room-temperature ferromagnetism in two-dimensional Fe(3)GeTe(2). *Nature* **563**, 94-99, doi:10.1038/s41586-018-0626-9 (2018).

[207] Xianqing Lin & Jun Ni. Layer-dependent intrinsic anomalous Hall effect in Fe3GeTe2. *Physical Review B* **100**, doi:10.1103/PhysRevB.100.085403 (2019).

[208] K. Kim, J. Seo, E. Lee, K. T. Ko, B. S. Kim, B. G. Jang, J. M. Ok, J. Lee, Y. J. Jo, W. Kang, J. H. Shim, C. Kim, H. W. Yeom, B. Il Min, B. J. Yang & J. S. Kim. Large anomalous Hall current induced by topological nodal lines in a ferromagnetic van der Waals semimetal. *Nat Mater* **17**, 794-799, doi:10.1038/s41563-018-0132-3 (2018).

[209] Hans-Jörg Deiseroth, Krasimir Aleksandrov, Christof Reiner, Lorenz Kienle & Reinhard K. Kremer. Fe3GeTe2 and Ni3GeTe2 – Two New Layered Transition-Metal Compounds: Crystal Structures, HRTEM Investigations, and Magnetic and Electrical Properties. *European Journal of Inorganic Chemistry* **2006**, 1561-1567, doi:10.1002/ejic.200501020 (2006).

[210] Jieyu Yi, Houlong Zhuang, Qiang Zou, Zhiming Wu, Guixin Cao, Siwei Tang, S. A. Calder, P. R. C. Kent, David Mandrus & Zheng Gai. Competing antiferromagnetism in a quasi-2D itinerant ferromagnet: Fe

GeTe

2. *2D Materials* **4**, doi:10.1088/2053-1583/4/1/011005 (2016).

[211] Yihao Wang, Cong Xian, Jian Wang, Bingjie Liu, Langsheng Ling, Lei Zhang, Liang Cao, Zhe Qu & Yimin Xiong. Anisotropic anomalous Hall effect in triangular itinerant

ferromagnet

Fe3GeTe2. *Physical Review B* **96**, doi:10.1103/PhysRevB.96.134428 (2017).

[212]  J. Ke, M. Yang, W. Xia, H. Zhu, C. Liu, R. Chen, C. Dong, W. Liu, M. Shi, Y. Guo & J. Wang. Magnetic and magneto-transport studies of two-dimensional ferromagnetic compound Fe(3)GeTe(2). *J Phys Condens Matter* **32**, 405805, doi:10.1088/1361-648X/ab9bc9 (2020).

[213]  Honglei Feng, Yong Li, Youguo Shi, Hong-Yi Xie, Yongqing Li & Yang Xu. Resistance anomaly and linear magnetoresistance in thin flakes of itinerant ferromagnet Fe3GeTe2. *Chinese Physics Letters* **39**, doi:10.1088/0256-307x/39/7/077501 (2022).

[214]  J. Xu, W. A. Phelan & C. L. Chien. Large Anomalous Nernst Effect in a van der Waals Ferromagnet Fe(3)GeTe(2). *Nano Lett* **19**, 8250-8254, doi:10.1021/acs.nanolett.9b03739 (2019).

[215]  Z. Fei, B. Huang, P. Malinowski, W. Wang, T. Song, J. Sanchez, W. Yao, D. Xiao, X. Zhu, A. F. May, W. Wu, D. H. Cobden, J. H. Chu & X. Xu. Two-dimensional itinerant ferromagnetism in atomically thin Fe(3)GeTe(2). *Nat Mater* **17**, 778-782, doi:10.1038/s41563-018-0149-7 (2018).

[216]  Q. Li, M. Yang, C. Gong, R. V. Chopdekar, A. T. N'Diaye, J. Turner, G. Chen, A. Scholl, P. Shafer, E. Arenholz, A. K. Schmid, S. Wang, K. Liu, N. Gao, A. S. Admasu, S. W. Cheong, C. Hwang, J. Li, F. Wang, X. Zhang & Z. Qiu. Patterning-Induced Ferromagnetism of Fe(3)GeTe(2) van der Waals Materials beyond Room Temperature. *Nano Lett* **18**, 5974-5980, doi:10.1021/acs.nanolett.8b02806 (2018).

[217]  C. Tan, J. Lee, S. G. Jung, T. Park, S. Albarakati, J. Partridge, M. R. Field, D. G. McCulloch, L. Wang & C. Lee. Hard magnetic properties in nanoflake van der Waals Fe(3)GeTe(2). *Nat Commun* **9**, 1554, doi:10.1038/s41467-018-04018-w (2018).

[218]  X. Wang, J. Tang, X. Xia, C. He, J. Zhang, Y. Liu, C. Wan, C. Fang, C. Guo, W. Yang, Y. Guang, X. Zhang, H. Xu, J. Wei, M. Liao, X. Lu, J. Feng, X. Li, Y. Peng, H. Wei, R. Yang, D. Shi, X. Zhang, Z. Han, Z. Zhang, G. Zhang, G. Yu & X. Han. Current-driven magnetization switching in a van der Waals ferromagnet Fe(3)GeTe(2). *Sci Adv* **5**, eaaw8904, doi:10.1126/sciadv.aaw8904 (2019).

[219]  S. Y. Park, D. S. Kim, Y. Liu, J. Hwang, Y. Kim, W. Kim, J. Y. Kim, C. Petrovic, C. Hwang, S. K. Mo, H. J. Kim, B. C. Min, H. C. Koo, J. Chang, C. Jang, J. W. Choi & H. Ryu. Controlling the Magnetic Anisotropy of the van der Waals Ferromagnet Fe(3)GeTe(2) through Hole Doping. *Nano Lett* **20**, 95-100, doi:10.1021/acs.nanolett.9b03316 (2020).

[220]  H. Wang, Y. Liu, P. Wu, W. Hou, Y. Jiang, X. Li, C. Pandey, D. Chen, Q. Yang, H. Wang, D. Wei, N. Lei, W. Kang, L. Wen, T. Nie, W. Zhao & K. L. Wang. Above Room-Temperature Ferromagnetism in Wafer-Scale Two-Dimensional van der Waals Fe(3)GeTe(2) Tailored by a Topological Insulator. *ACS Nano* **14**, 10045-10053, doi:10.1021/acsnano.0c03152 (2020).

[221]  In Kee Park, Cheng Gong, Kyoo Kim & Geunsik Lee. Controlling interlayer magnetic coupling in the two-dimensional magnet

Fe3GeTe2. *Physical Review B* **105**, doi:10.1103/PhysRevB.105.014406 (2022).

[222]  H. P. Wang, D. S. Wu, Y. G. Shi & N. L. Wang. Anisotropic transport and optical spectroscopy study on antiferromagnetic triangular latticeEuCd2As2: An interplay between magnetism and charge transport properties. *Physical Review B* **94**, doi:10.1103/PhysRevB.94.045112 (2016).

[223] M. C. Rahn, J. R. Soh, S. Francoual, L. S. I. Veiga, J. Strempfer, J. Mardegan, D. Y. Yan, Y. F. Guo, Y. G. Shi & A. T. Boothroyd. Coupling of magnetic order and charge transport in the candidate Dirac semimetal EuCd2As2. *Physical Review B* **97**, doi:10.1103/PhysRevB.97.214422 (2018).

[224] K. M. Taddei, L. Yin, L. D. Sanjeewa, Y. Li, J. Xing, C. dela Cruz, D. Phelan, A. S. Sefat & D. S. Parker. Single pair of Weyl nodes in the spin-canted structure of EuCd2As2. *Physical Review B* **105**, doi:10.1103/PhysRevB.105.L140401 (2022).

[225] J. Ma, H. Wang, S. Nie, C. Yi, Y. Xu, H. Li, J. Jandke, W. Wulfhekel, Y. Huang, D. West, P. Richard, A. Chikina, V. N. Strocov, J. Mesot, H. Weng, S. Zhang, Y. Shi, T. Qian, M. Shi & H. Ding. Emergence of Nontrivial Low-Energy Dirac Fermions in Antiferromagnetic EuCd(2)As(2). *Adv Mater* **32**, e1907565, doi:10.1002/adma.201907565 (2020).

[226] Xiangyu Cao, Jie-Xiang Yu, Pengliang Leng, Changjiang Yi, Xiaoyang Chen, Yunkun Yang, Shanshan Liu, Lingyao Kong, Zihan Li, Xiang Dong, Youguo Shi, Manuel Bibes, Rui Peng, Jiadong Zang & Faxian Xiu. Giant nonlinear anomalous Hall effect induced by spin-dependent band structure evolution. *Physical Review Research* **4**, doi:10.1103/PhysRevResearch.4.023100 (2022).

[227] Inga Schellenberg, Ulrike Pfannenschmidt, Matthias Eul, Christian Schwickert & Rainer Pöttgen. A 121Sb and 151Eu Mössbauer Spectroscopic Investigation of EuCd2X2 (X = P, As, Sb) and YbCd2Sb2. *Zeitschrift für anorganische und allgemeine Chemie* **637**, 1863-1870, doi:10.1002/zaac.201100179 (2011).

[228] Lin-Lin Wang, Na Hyun Jo, Brinda Kuthanazhi, Yun Wu, Robert J. McQueeney, Adam Kaminski & Paul C. Canfield. Single pair of Weyl fermions in the half-metallic semimetal EuCd2As2. *Physical Review B* **99**, doi:10.1103/PhysRevB.99.245147 (2019).

[229] J. R. Soh, C. Donnerer, K. M. Hughes, E. Schierle, E. Weschke, D. Prabhakaran & A. T. Boothroyd. Magnetic and electronic structure of the layered rare-earth pnictide EuCd2Sb2. *Physical Review B* **98**, doi:10.1103/PhysRevB.98.064419 (2018).

[230] Jyoti Krishna, T. Nautiyal & T. Maitra. First-principles study of electronic structure, transport, and optical properties of EuCd2As2. *Physical Review B* **98**, doi:10.1103/PhysRevB.98.125110 (2018).

[231] Yingkai Sun, Yong Li, Shuaishuai Li, Changjiang Yi, Hanbin Deng, Xin Du, Limin Liu, Changjiang Zhu, Yuan Li, Zheng Wang, Hanqing Mao, Youguo Shi & Rui Wu. Experimental evidence for field-induced metamagnetic transition of EuCd2As2. *Journal of Rare Earths* **40**, 1606-1610, doi:10.1016/j.jre.2021.08.002 (2022).

[232] Guiyuan Hua, Simin Nie, Zhida Song, Rui Yu, Gang Xu & Kailun Yao. Dirac semimetal in type-IV magnetic space groups. *Physical Review B* **98**, doi:10.1103/PhysRevB.98.201116 (2018).

[233] Frank Schindler, Ashley M. Cook, Maia G. Vergniory, Zhijun Wang, Stuart S. P. Parkin, B. Andrei Bernevig & Titus Neupert. Higher-order topological insulators. *Science Advances* **4**, eaat0346, doi:10.1126/sciadv.aat0346 (2018).

[234] J. R. Soh, F. de Juan, M. G. Vergniory, N. B. M. Schröter, M. C. Rahn, D. Y. Yan, J. Jiang, M. Bristow, P. A. Reiss, J. N. Blandy, Y. F. Guo, Y. G. Shi, T. K. Kim, A. McCollam, S. H. Simon, Y. Chen, A. I. Coldea & A. T. Boothroyd. Ideal Weyl semimetal induced by magnetic exchange. *Physical Review B* **100**, doi:10.1103/PhysRevB.100.201102 (2019).

[235] L. A. Fenner, A. A. Dee & A. S. Wills. Non-collinearity and spin frustration in the itinerant

kagome ferromagnet Fe3Sn2. *Journal of Physics: Condensed Matter* **21**, 452202, doi:10.1088/0953-8984/21/45/452202 (2009).

[236] L. Ye, M. Kang, J. Liu, F. von Cube, C. R. Wicker, T. Suzuki, C. Jozwiak, A. Bostwick, E. Rotenberg, D. C. Bell, L. Fu, R. Comin & J. G. Checkelsky. Massive Dirac fermions in a ferromagnetic kagome metal. *Nature* **555**, 638-642, doi:10.1038/nature25987 (2018).

[237] B Malaman, B Roques, A Courtois & J Protas. Structure cristalline du stannure de fer Fe3Sn2. *Acta Crystallographica Section B: Structural Crystallography and Crystal Chemistry* **32**, 1348-1351 (1976).

[238] G. Le Caer, B. Malaman & B. Roques. Mossbauer effect study of Fe3Sn2. *Journal of Physics F: Metal Physics* **8**, 323, doi:10.1088/0305-4608/8/2/018 (1978).

[239] B. Malaman, D. Fruchart & G. Le Caer. Magnetic properties of Fe3Sn2. II. Neutron diffraction study (and Mossbauer effect). *Journal of Physics F: Metal Physics* **8**, 2389, doi:10.1088/0305-4608/8/11/022 (1978).

[240] G. Le Caer, B. Malaman, L. Haggstrom & T. Ericsson. Magnetic properties of Fe3Sn2. III. A 119Sn Mossbauer study. *Journal of Physics F: Metal Physics* **9**, 1905, doi:10.1088/0305-4608/9/9/020 (1979).

[241] Z. Lin, J. H. Choi, Q. Zhang, W. Qin, S. Yi, P. Wang, L. Li, Y. Wang, H. Zhang, Z. Sun, L. Wei, S. Zhang, T. Guo, Q. Lu, J. H. Cho, C. Zeng & Z. Zhang. Flatbands and Emergent Ferromagnetic Ordering in Fe3Sn2 Kagome Lattices. *Phys Rev Lett* **121**, 096401, doi:10.1103/PhysRevLett.121.096401 (2018).

[242] J. X. Yin, S. S. Zhang, H. Li, K. Jiang, G. Chang, B. Zhang, B. Lian, C. Xiang, I. Belopolski, H. Zheng, T. A. Cochran, S. Y. Xu, G. Bian, K. Liu, T. R. Chang, H. Lin, Z. Y. Lu, Z. Wang, S. Jia, W. Wang & M. Z. Hasan. Giant and anisotropic many-body spin-orbit tunability in a strongly correlated kagome magnet. *Nature* **562**, 91-95, doi:10.1038/s41586-018-0502-7 (2018).

[243] Qi Wang, Shanshan Sun, Xiao Zhang, Fei Pang & Hechang Lei. Anomalous Hall effect in a ferromagneticFe3Sn2single crystal with a geometrically frustrated Fe bilayer kagome lattice. *Physical Review B* **94**, doi:10.1103/PhysRevB.94.075135 (2016).

[244] Zhi-Peng Hou, Bei Ding, Hang Li, Gui-Zhou Xu, Wen-Hong Wang & Guang-Heng Wu. Observation of new-type magnetic skymrions with extremerely high temperature stability and fabrication of skyrmion-based race-track memory device. *Acta Physica Sinica* **67**, doi:10.7498/aps.67.20180419 (2018).

[245] Hang Li, Bei Ding, Jie Chen, Zefang Li, Zhipeng Hou, Enke Liu, Hongwei Zhang, Xuekui Xi, Guangheng Wu & Wenhong Wang. Large topological Hall effect in a geometrically frustrated kagome magnet Fe3Sn2. *Applied Physics Letters* **114**, doi:10.1063/1.5088173 (2019).

[246] Christopher D. O'Neill, Andrew S. Wills & Andrew D. Huxley. Possible topological contribution to the anomalous Hall effect of the noncollinear ferromagnet
Fe3Sn2. *Physical Review B* **100**, doi:10.1103/PhysRevB.100.174420 (2019).

[247] Qi Wang, Qiangwei Yin & Hechang Lei. Giant topological Hall effect of ferromagnetic kagome metal Fe3Sn2*. *Chinese Physics B* **29**, doi:10.1088/1674-1056/ab5fbc (2020).

[248] Z. Hou, W. Ren, B. Ding, G. Xu, Y. Wang, B. Yang, Q. Zhang, Y. Zhang, E. Liu, F. Xu, W. Wang, G. Wu, X. Zhang, B. Shen & Z. Zhang. Observation of Various and Spontaneous Magnetic Skyrmionic Bubbles at Room Temperature in a Frustrated Kagome Magnet with

Uniaxial Magnetic Anisotropy. *Adv Mater* **29**, doi:10.1002/adma.201701144 (2017).

[249] Z. Hou, Q. Zhang, G. Xu, C. Gong, B. Ding, Y. Wang, H. Li, E. Liu, F. Xu, H. Zhang, Y. Yao, G. Wu, X. X. Zhang & W. Wang. Creation of Single Chain of Nanoscale Skyrmion Bubbles with Record-High Temperature Stability in a Geometrically Confined Nanostripe. *Nano Lett* **18**, 1274-1279, doi:10.1021/acs.nanolett.7b04900 (2018).

[250] Lingling Gao, Shiwei Shen, Qi Wang, Wujun Shi, Yi Zhao, Changhua Li, Weizheng Cao, Cuiying Pei, Jun-Yi Ge, Gang Li, Jun Li, Yulin Chen, Shichao Yan & Yanpeng Qi. Anomalous Hall effect in ferrimagnetic metal RMn6Sn6 (R = Tb, Dy, Ho) with clean Mn kagome lattice. *Applied Physics Letters* **119**, doi:10.1063/5.0061260 (2021).

[251] J. X. Yin, W. Ma, T. A. Cochran, X. Xu, S. S. Zhang, H. J. Tien, N. Shumiya, G. Cheng, K. Jiang, B. Lian, Z. Song, G. Chang, I. Belopolski, D. Multer, M. Litskevich, Z. J. Cheng, X. P. Yang, B. Swidler, H. Zhou, H. Lin, T. Neupert, Z. Wang, N. Yao, T. R. Chang, S. Jia & M. Zahid Hasan. Quantum-limit Chern topological magnetism in TbMn(6)Sn(6). *Nature* **583**, 533-536, doi:10.1038/s41586-020-2482-7 (2020).

[252] Dong Chen, Congcong Le, Chenguang Fu, Haicheng Lin, Walter Schnelle, Yan Sun & Claudia Felser. Large anomalous Hall effect in the kagome ferromagnet LiMn6Sn6. *Physical Review B* **103**, doi:10.1103/PhysRevB.103.144410 (2021).

[253] B. Chafik El Idrissi, G. Venturini & B. Malaman. Crystal structures of RFe6Sn6 (R = Sc, Y, Gd-Tm, Lu) rare-earth iron stannides. *Materials Research Bulletin* **26**, 1331-1338, doi:https://doi.org/10.1016/0025-5408(91)90149-G (1991).

[254] G. Venturini, B. Chafik El Idrissi & B. Malaman. Magnetic properties of RMn6Sn6 (R = Sc, Y, Gd—Tm, Lu) compounds with HfFe6Ge6 type structure. *Journal of Magnetism and Magnetic Materials* **94**, 35-42, doi:https://doi.org/10.1016/0304-8853(91)90108-M (1991).

[255] Nirmal J. Ghimire, Rebecca L. Dally, L. Poudel, D. C. Jones, D. Michel, N. Thapa Magar, M. Bleuel, Michael A. McGuire, J. S. Jiang, J. F. Mitchell, Jeffrey W. Lynn & I. I. Mazin. Competing magnetic phases and fluctuation-driven scalar spin chirality in the kagome metal YMn6Sn6. *Science Advances* **6**, eabe2680, doi:10.1126/sciadv.abe2680 (2020).

[256] W. Ma, X. Xu, J. X. Yin, H. Yang, H. Zhou, Z. J. Cheng, Y. Huang, Z. Qu, F. Wang, M. Z. Hasan & S. Jia. Rare Earth Engineering in RMn_6Sn_6 (R=Gd-Tm, Lu) Topological Kagome Magnets. *Phys Rev Lett* **126**, 246602, doi:10.1103/PhysRevLett.126.246602 (2021).

[257] M. Li, Q. Wang, G. Wang, Z. Yuan, W. Song, R. Lou, Z. Liu, Y. Huang, Z. Liu, H. Lei, Z. Yin & S. Wang. Dirac cone, flat band and saddle point in kagome magnet YMn(6)Sn(6). *Nat Commun* **12**, 3129, doi:10.1038/s41467-021-23536-8 (2021).

[258] X. Gu, C. Chen, W. S. Wei, L. L. Gao, J. Y. Liu, X. Du, D. Pei, J. S. Zhou, R. Z. Xu, Z. X. Yin, W. X. Zhao, Y. D. Li, C. Jozwiak, A. Bostwick, E. Rotenberg, D. Backes, L. S. I. Veiga, S. Dhesi, T. Hesjedal, G. van der Laan, H. F. Du, W. J. Jiang, Y. P. Qi, G. Li, W. J. Shi, Z. K. Liu, Y. L. Chen & L. X. Yang. Robust kagome electronic structure in the topological quantum magnets XMn6Sn6 (X=Dy,Tb,Gd,Y). *Physical Review B* **105**, doi:10.1103/PhysRevB.105.155108 (2022).

[259] S. Roychowdhury, A. M. Ochs, S. N. Guin, K. Samanta, J. Noky, C. Shekhar, M. G. Vergniory, J. E. Goldberger & C. Felser. Large Room Temperature Anomalous Transverse Thermoelectric Effect in Kagome Antiferromagnet YMn(6) Sn(6). *Adv Mater* **34**, e2201350, doi:10.1002/adma.202201350 (2022).

[260] Gyanendra Dhakal, Fairoja Cheenicode Kabeer, Arjun K. Pathak, Firoza Kabir, Narayan Poudel, Randall Filippone, Jacob Casey, Anup Pradhan Sakhya, Sabin Regmi, Christopher Sims, Klauss Dimitri, Pietro Manfrinetti, Krzysztof Gofryk, Peter M. Oppeneer & Madhab Neupane. Anisotropically large anomalous and topological Hall effect in a kagome magnet. *Physical Review B* **104**, doi:10.1103/PhysRevB.104.L161115 (2021).

[261] Qi Wang, Kelly J. Neubauer, Chunruo Duan, Qiangwei Yin, Satoru Fujitsu, Hideo Hosono, Feng Ye, Rui Zhang, Songxue Chi, Kathryn Krycka, Hechang Lei & Pengcheng Dai. Field-induced topological Hall effect and double-fan spin structure with a c-axis component in the metallic kagome antiferromagnetic compound YMn6Sn6. *Physical Review B* **103**, doi:10.1103/PhysRevB.103.014416 (2021).

[262] Firoza Kabir, Randall Filippone, Gyanendra Dhakal, Y. Lee, Narayan Poudel, Jacob Casey, Anup Pradhan Sakhya, Sabin Regmi, Robert Smith, Pietro Manfrinetti, Liqin Ke, Krzysztof Gofryk, Madhab Neupane & Arjun K. Pathak. Unusual magnetic and transport properties in HoMn6Sn6 kagome magnet. *Physical Review Materials* **6**, doi:10.1103/PhysRevMaterials.6.064404 (2022).

[263] Jeonghun Lee & Eundeok Mun. Anisotropic magnetic property of single crystals RV6Sn6 (R=Y, Gd−Tm, Lu). *Physical Review Materials* **6**, doi:10.1103/PhysRevMaterials.6.083401 (2022).

[264] S. Peng, Y. Han, G. Pokharel, J. Shen, Z. Li, M. Hashimoto, D. Lu, B. R. Ortiz, Y. Luo, H. Li, M. Guo, B. Wang, S. Cui, Z. Sun, Z. Qiao, S. D. Wilson & J. He. Realizing Kagome Band Structure in Two-Dimensional Kagome Surface States of RV_6Sn_6 (R=Gd, Ho). *Phys Rev Lett* **127**, 266401, doi:10.1103/PhysRevLett.127.266401 (2021).

[265] Yong Hu, Xianxin Wu, Yongqi Yang, Shunye Gao, Nicholas C. Plumb, Andreas P. Schnyder, Weiwei Xie, Junzhang Ma & Ming Shi. Tunable topological Dirac surface states and van Hove singularities in kagome metal GdV6Sn6. *Science Advances* **8**, eadd2024, doi:10.1126/sciadv.add2024 (2022).

[266] E. Cheng, W. Xia, X. Shi, H. Fang, C. Wang, C. Xi, S. Xu, D. C. Peets, L. Wang, H. Su, L. Pi, W. Ren, X. Wang, N. Yu, Y. Chen, W. Zhao, Z. Liu, Y. Guo & S. Li. Magnetism-induced topological transition in EuAs(3). *Nat Commun* **12**, 6970, doi:10.1038/s41467-021-26482-7 (2021).

[267] W. Bauhofer, M. Wittmann & H. G. v Schnering. Structure, electrical and magnetic properties of CaAs3, SrAs3, BaAs3 and EuAs3. *Journal of Physics and Chemistry of Solids* **42**, 687-695, doi:10.1016/0022-3697(81)90122-0 (1981).

[268] T. Chattopadhyay, H. G. v. Schnering & P. J. Brown. Neutron diffraction study of the magnetic ordering in EuAs3. *Journal of Magnetism and Magnetic Materials* **28**, 247-249, doi:10.1016/0304-8853(82)90056-7 (1982).

[269] T. Chattopadhyay & P. J. Brown. Field-induced transverse-sine-wave-to-longitudinal-sine-wave transition in EuAs3. *Phys Rev B Condens Matter* **38**, 795-797, doi:10.1103/physrevb.38.795 (1988).

[270] Tapan Chatterji, K. D. Liß, T. Tschentscher, B. Janossy, J. Strempfer & T. Brückel. High-energy non-resonant X-ray magnetic scattering from EuAs3. *Solid State Communications* **131**, 713-717, doi:10.1016/j.ssc.2004.06.026 (2004).

[271] Tapan Chatterji & Wolfgang Henggeler. μSR investigation of the magnetic ordering in EuAs3. *Solid State Communications* **132**, 617-622, doi:10.1016/j.ssc.2004.08.033 (2004).

[272] W. Bauhofer & K. A. McEwen. Anisotropic magnetoresistance of the semimetallic antiferromagnet EuAs3. *Phys Rev B Condens Matter* **43**, 13450-13455, doi:10.1103/physrevb.43.13450 (1991).

[273] J. Gooth, B. Bradlyn, S. Honnali, C. Schindler, N. Kumar, J. Noky, Y. Qi, C. Shekhar, Y. Sun, Z. Wang, B. A. Bernevig & C. Felser. Axionic charge-density wave in the Weyl semimetal (TaSe4)2I. *Nature* **575**, 315-319, doi:10.1038/s41586-019-1630-4 (2019).

[274] Libor Šmejkal, Allan H. MacDonald, Jairo Sinova, Satoru Nakatsuji & Tomas Jungwirth. Anomalous Hall antiferromagnets. *Nature Reviews Materials* **7**, 482-496, doi:10.1038/s41578-022-00430-3 (2022).